\documentclass[journal]{IEEEtran}
\usepackage{xspace}
\usepackage{xcolor}
\usepackage{colortbl}
\usepackage{cite}
\usepackage{url}

\usepackage{todonotes}
\usepackage{graphicx}
\usepackage{diagbox}
\usepackage{amsthm}
\usepackage{amsmath}
\usepackage{amssymb}
\usepackage{cleveref}
\usepackage{array}
\usepackage{setspace}
\usepackage{cases}
\usepackage{makecell}
\usepackage{multirow}
\usepackage{listings}
\usepackage{extarrows}
\usepackage{float}
\usepackage{subfigure}

\usepackage{algorithm}
\usepackage{algorithmic}

\usepackage{pifont}

\newtheorem{myDef}{Definition}
\def\oursystem{MalFlows\xspace}
\def\ie{\textit{i.e.}\xspace}
\def\etc{\textit{etc.}\xspace}
\def\eg{\textit{e.g.}\xspace}

\ifCLASSINFOpdf
\else
\fi
\begin{document}
%
% paper title
% Titles are generally capitalized except for words such as a, an, and, as,
% at, but, by, for, in, nor, of, on, or, the, to and up, which are usually
% not capitalized unless they are the first or last word of the title.
% Linebreaks \\ can be used within to get better formatting as desired.
% Do not put math or special symbols in the title.
\title{\oursystem: Context-aware Fusion of Heterogeneous Flow Semantics for Android Malware Detection}
%
%
% author names and IEEE memberships
% note positions of commas and nonbreaking spaces ( ~ ) LaTeX will not break
% a structure at a ~ so this keeps an author's name from being broken across
% two lines.
% use \thanks{} to gain access to the first footnote area
% a separate \thanks must be used for each paragraph as LaTeX2e's \thanks
% was not built to handle multiple paragraphs
%

\author{Zhaoyi Meng, Fenglei Xu, Wenxiang Zhao, Wansen Wang, Wenchao Huang, Jie Cui, Hong Zhong, and Yan Xiong% <-this % stops a space
\thanks{Zhaoyi Meng, Fenglei Xu, Wansen Wang, Jie Cui, Hong Zhong are with the School of Computer Science and Technology, Anhui University, Hefei, 230039, China. Wenxiang Zhao, Wenchao Huang, Yan Xiong are with the School of Computer Science and Technology, University of Science and Technology of China, Hefei 230027, China. Corresponding authors: Wansen Wang, Wenchao Huang. (e-mail: zymeng@ahu.edu.cn; 23762@ahu.edu.cn; huangwc@ustc.edu.cn)}}
\maketitle

% For peer review papers, you can put extra information on the cover
% page as needed:
% \ifCLASSOPTIONpeerreview
% \begin{center} \bfseries EDICS Category: 3-BBND \end{center}
% \fi
%
% For peerreview papers, this IEEEtran command inserts a page break and
% creates the second title. It will be ignored for other modes.
\IEEEpeerreviewmaketitle

%!TEX root = bare_jrnl.tex

\begin{abstract}
Static analysis, a fundamental technique in Android app examination, enables the extraction of control flows, data flows, and inter-component communications (ICCs), all of which are essential for malware detection.
However, existing methods struggle to leverage the semantic complementarity across different types of flows for representing program behaviors, and their context-unaware nature further hinders the accuracy of cross-flow semantic integration.
We propose and implement \textit{\oursystem}, a novel technique that achieves context-aware fusion of heterogeneous flow semantics for Android malware detection.
Our goal is to leverage complementary strengths of the three types of flow-related information for precise app profiling.
We adopt a heterogeneous information network (HIN) to model the rich semantics across these program flows.
We further propose \textit{flow2vec}, a context-aware HIN embedding technique that distinguishes the semantics of HIN entities as needed based on contextual constraints across different flows and learns accurate app representations through the joint use of multiple meta-paths.
The representations are finally fed into a channel-attention-based deep neural network for malware classification.
To the best of our knowledge, this is the first study to comprehensively aggregate the strengths of diverse flow-related information for assessing maliciousness within apps.
We evaluate \oursystem on a large-scale dataset comprising over 20 million flow instances extracted from more than 31,000 real-world apps.
Experimental results demonstrate that \oursystem outperforms representative baselines in Android malware detection, and meanwhile, validate the effectiveness of \textit{flow2vec} in accurately learning app representations from the HIN constructed over the heterogeneous flows.
\end{abstract}

% Note that keywords are not normally used for peerreview papers.
% \begin{IEEEkeywords}
% Mobile security, Android malware detection, heterogeneous information network, multi-view fusion.
% \end{IEEEkeywords}
%!TEX root = bare_jrnl.tex

\section{Introduction}

\IEEEPARstart{W}{ith} the highest market share worldwide on mobile operating systems~\cite{MarketShare2024} and the openness of development, the Android platform is targeted by various malware and other potentially harmful apps.
Such apps steal users' privacy, abuse SMS/CALL, subscribe to premium services silently, \textit{etc}. 
Meanwhile, the characteristics of Android malware have been evolving over years~\cite{wang2022malradar}.
The circumstance calls for effective and reliable detection techniques in the Android ecosystem.

Static analysis techniques commonly extract control flows, data flows, and inter-component communications (ICCs) as fundamental products for the examination of inherent semantics within app behaviors to uncover hidden malicious activities~\cite{yang2015static,DBLP:conf/pldi/ArztRFBBKTOM14,DBLP:conf/icse/0029BBKTARBOM15}.
Control flows determine the execution orders of program statements, data flows trace the paths that data takes through the system from input to output, and ICCs manage the interactions between different app components.
The three types of flow-related semantic information are inherently associated with various APIs, involving triggering conditions, data transmissions and communication modes, which serve as crucial indicators for assessing maliciousness of app behaviors.
Thus, effectively leveraging these flow-related semantics is essential for achieving precise malware detection.

Existing malware detection schemes have not fully leveraged the synergistic capabilities of heterogeneous flow semantics.
In particular, flow-based approaches usually utilize only a part of the information within apps in isolation.
For instance, MUDFLOW~\cite{DBLP:conf/icse/AvdiienkoKGZARB15}, TriFlow~\cite{mirzaei2017triflow}, and Complex-Flows~\cite{shen2018android} detected malware by learning usage characteristics of sensitive data flows.
ICCDetector~\cite{xu2016iccdetector} focused on ICC-based features for detection.
VAHunt~\cite{shi2020vahunt} and Difuzer~\cite{samhi2022difuzer} respectively identified malicious virtualization-based apps and logic bombs, using control-flow analysis and taint tracking in separate steps.
These approaches can result in incomplete analysis and potentially miss certain malicious behaviors.
For non-flow-based methods, the lack of the fine-grained program clues makes it challenging to assess intentions of app behaviors.
This limitation applies to conventional~\cite{arora2019permpair}, machine learning(ML)-based~\cite{arp2014drebin,he2023msdroid,kim2018multimodal,meng2025detecting}, and large language model(LLM)-based~\cite{zhao2025apppoet} techniques.
 
\begin{table*}[t]
\caption{Statistical results of top-10 key elements from the views of control flows, data flows, and ICCs on our whole dataset}\label{stats}
\centering
\resizebox{0.96\linewidth}{!}{
\begin{tabular}{|llllllll|lllll|}
\hline
\multicolumn{8}{|c|}{\cellcolor[HTML]{E6E6E6}\textbf{Triggering Conditions}}                                                                                       & \multicolumn{5}{c|}{\cellcolor[HTML]{E6E6E6}\textbf{Guarded APIs (Simplified)}}                                                                          \\ \hline
\multicolumn{2}{|l|}{\textbf{Benignware}}    & \multicolumn{2}{l|}{\textbf{\#}} & \multicolumn{2}{l|}{\textbf{Malware}}        & \multicolumn{2}{l|}{\textbf{\#}} & \multicolumn{1}{l|}{\textbf{Benignware}}     & \multicolumn{2}{l|}{\textbf{\#}} & \multicolumn{1}{l|}{\textbf{Malware}}                   & \textbf{\#} \\ \hline
\multicolumn{2}{|l|}{NO\_CATEGORY}           & \multicolumn{2}{l|}{72,828}       & \multicolumn{2}{l|}{NO\_CATEGORY}            & \multicolumn{2}{l|}{149,114}      & \multicolumn{1}{l|}{res.AssetManager.open}   & \multicolumn{2}{l|}{4,035}        & \multicolumn{1}{l|}{res.AssetManager.open}              & 46,456       \\ \hline
\multicolumn{2}{|l|}{NETWORK\_INFORMATION}   & \multicolumn{2}{l|}{6,690}        & \multicolumn{2}{l|}{NETWORK\_INFORMATION}    & \multicolumn{2}{l|}{43,512}       & \multicolumn{1}{l|}{HashMap.put}             & \multicolumn{2}{l|}{3,920}        & \multicolumn{1}{l|}{res.Resources.getAssets}            & 46,167       \\ \hline
\multicolumn{2}{|l|}{DATABASE\_INFORMATION}  & \multicolumn{2}{l|}{2,908}        & \multicolumn{2}{l|}{DATABASE\_INFORMATION}   & \multicolumn{2}{l|}{824}         & \multicolumn{1}{l|}{String.startsWith}       & \multicolumn{2}{l|}{3,474}        & \multicolumn{1}{l|}{FileOutputStream. write}             & 27,411       \\ \hline
\multicolumn{2}{|l|}{LOCATION\_INFORMATION}  & \multicolumn{2}{l|}{819}         & \multicolumn{2}{l|}{UNIQUE\_IDENTIFIER}      & \multicolumn{2}{l|}{620}         & \multicolumn{1}{l|}{res.Resources.getAssets} & \multicolumn{2}{l|}{3,419}        & \multicolumn{1}{l|}{File.getParentFile}                 & 18,802       \\ \hline
\multicolumn{2}{|l|}{BLUETOOTH\_INFORMATION} & \multicolumn{2}{l|}{241}         & \multicolumn{2}{l|}{LOCATION\_INFORMATION}   & \multicolumn{2}{l|}{301}         & \multicolumn{1}{l|}{FileOutputStream.write}  & \multicolumn{2}{l|}{3,317}        & \multicolumn{1}{l|}{File.getAbsolutePath}               & 13,612       \\ \hline
\multicolumn{2}{|l|}{CALENDAR\_INFORMATION}  & \multicolumn{2}{l|}{181}         & \multicolumn{2}{l|}{BLUETOOTH\_INFORMATION}  & \multicolumn{2}{l|}{59}          & \multicolumn{1}{l|}{File.getPath}            & \multicolumn{2}{l|}{2,620}        & \multicolumn{1}{l|}{File.getCanonicalPath}              & 11,543       \\ \hline
\multicolumn{2}{|l|}{UNIQUE\_IDENTIFIER}     & \multicolumn{2}{l|}{169}         & \multicolumn{2}{l|}{CALENDAR\_INFORMATION}   & \multicolumn{2}{l|}{33}          & \multicolumn{1}{l|}{Activity.getIntent}      & \multicolumn{2}{l|}{2,256}        & \multicolumn{1}{l|}{URLConnection.setRequestProperty}   & 1,721        \\ \hline
\multicolumn{2}{|l|}{FILE\_INFORMATION}      & \multicolumn{2}{l|}{90}          & \multicolumn{2}{l|}{ACCOUNT\_INFORMATION}    & \multicolumn{2}{l|}{4}           & \multicolumn{1}{l|}{File.getAbsolutePath}    & \multicolumn{2}{l|}{2,197}        & \multicolumn{1}{l|}{URLConnection.getInputStream}       & 893         \\ \hline
\multicolumn{2}{|l|}{NFC}                    & \multicolumn{2}{l|}{62}          & \multicolumn{2}{l|}{NFC}                     & \multicolumn{2}{l|}{1}           & \multicolumn{1}{l|}{File.getParentFile}      & \multicolumn{2}{l|}{2,062}        & \multicolumn{1}{l|}{HttpURLConnection.setRequestMethod} & 856         \\ \hline
\multicolumn{2}{|l|}{ACCOUNT\_INFORMATION}   & \multicolumn{2}{l|}{2}           & \multicolumn{2}{l|}{FILE\_INFORMATION}       & \multicolumn{2}{l|}{0}           & \multicolumn{1}{l|}{Class.getName}           & \multicolumn{2}{l|}{1,772}        & \multicolumn{1}{l|}{Camera.setFlashMode}                & 467         \\ \hline
\multicolumn{8}{|c|}{\cellcolor[HTML]{E6E6E6}\textbf{Source APIs (Simplified)}}                                                                                    & \multicolumn{5}{c|}{\cellcolor[HTML]{E6E6E6}\textbf{Sink APIs (Simplified)}}                                                                             \\ \hline
\multicolumn{2}{|l|}{\textbf{Benignware}}    & \multicolumn{2}{l|}{\textbf{\#}} & \multicolumn{2}{l|}{\textbf{Malware}}        & \multicolumn{2}{l|}{\textbf{\#}}          & \multicolumn{1}{l|}{\textbf{Benignware}}     & \multicolumn{2}{l|}{\textbf{\#}} & \multicolumn{1}{l|}{\textbf{Malware}}                   & \textbf{\#} \\ \hline
\multicolumn{2}{|l|}{Class.getName}          & \multicolumn{2}{l|}{1,140,128}     & \multicolumn{2}{l|}{Class.getDeclaredMethod} & \multicolumn{2}{l|}{640,052}      & \multicolumn{1}{l|}{HashMap.put}             & \multicolumn{2}{l|}{2,234,284}     & \multicolumn{1}{l|}{HashMap.put}                        & 1,663,037     \\ \hline
\multicolumn{2}{|l|}{HashMap.get}            & \multicolumn{2}{l|}{554,018}      & \multicolumn{2}{l|}{Class.getName}           & \multicolumn{2}{l|}{481456}      & \multicolumn{1}{l|}{String.substring}        & \multicolumn{2}{l|}{1,950,565}     & \multicolumn{1}{l|}{String.substring}                   & 710,207      \\ \hline
\multicolumn{2}{|l|}{reflect.Field.get}      & \multicolumn{2}{l|}{534,559}      & \multicolumn{2}{l|}{HashMap.get}             & \multicolumn{2}{l|}{235,556}      & \multicolumn{1}{l|}{String.startsWith}       & \multicolumn{2}{l|}{1,005,406}     & \multicolumn{1}{l|}{String.startsWith}                  & 477,782      \\ \hline
\multicolumn{2}{|l|}{Hashtable.get}          & \multicolumn{2}{l|}{297,028}      & \multicolumn{2}{l|}{GregorianCalendar.get}   & \multicolumn{2}{l|}{212,738}      & \multicolumn{1}{l|}{reflect.Field.set}       & \multicolumn{2}{l|}{550,355}      & \multicolumn{1}{l|}{JSONObject.put}                     & 193,678      \\ \hline
\multicolumn{2}{|l|}{Class.getSimpleName}    & \multicolumn{2}{l|}{285,725}      & \multicolumn{2}{l|}{reflect.Field.get}       & \multicolumn{2}{l|}{201,261}      & \multicolumn{1}{l|}{URL.openConnection}      & \multicolumn{2}{l|}{382,903}      & \multicolumn{1}{l|}{reflect.Field.set}                  & 168,789      \\ \hline
\multicolumn{2}{|l|}{ArrayList.get}          & \multicolumn{2}{l|}{275,310}      & \multicolumn{2}{l|}{ArrayList.get}           & \multicolumn{2}{l|}{198,304}      & \multicolumn{1}{l|}{Log.d}                   & \multicolumn{2}{l|}{317,969}      & \multicolumn{1}{l|}{URL.openConnection}                 & 130,237      \\ \hline
\multicolumn{2}{|l|}{Array.newInstance}      & \multicolumn{2}{l|}{255,319}      & \multicolumn{2}{l|}{System.getProperty}      & \multicolumn{2}{l|}{103,814}      & \multicolumn{1}{l|}{Log.v}                   & \multicolumn{2}{l|}{265,275}      & \multicolumn{1}{l|}{Camera.setPreviewSize}              & 82,974       \\ \hline
\multicolumn{2}{|l|}{System.getProperty}     & \multicolumn{2}{l|}{250,467}      & \multicolumn{2}{l|}{SQLiteDatabase.query}    & \multicolumn{2}{l|}{54,504}       & \multicolumn{1}{l|}{StringBuffer.setLength}  & \multicolumn{2}{l|}{216,059}      & \multicolumn{1}{l|}{FileOutputStream.write}             & 65,697       \\ \hline
\multicolumn{2}{|l|}{ThreadLocal.get}        & \multicolumn{2}{l|}{247,968}      & \multicolumn{2}{l|}{reflect.Array.get}       & \multicolumn{2}{l|}{53,393}       & \multicolumn{1}{l|}{JSONObject.put}          & \multicolumn{2}{l|}{168,076}      & \multicolumn{1}{l|}{Log.w}                              & 65,323       \\ \hline
\multicolumn{2}{|l|}{Class.getMethod}        & \multicolumn{2}{l|}{178,051}      & \multicolumn{2}{l|}{File.getPath}            & \multicolumn{2}{l|}{51,140}       & \multicolumn{1}{l|}{ThreadLocal.set}         & \multicolumn{2}{l|}{158,014}      & \multicolumn{1}{l|}{StringBuffer.setLength}             & 61,138       \\ \hline
\multicolumn{8}{|c|}{\cellcolor[HTML]{E6E6E6}\textbf{Components}}                                                                                                  & \multicolumn{5}{c|}{\cellcolor[HTML]{E6E6E6}\textbf{Actions of Intents (Simplified)}}                                                                               \\ \hline
\multicolumn{2}{|l|}{\textbf{Benignware}}    & \multicolumn{2}{l|}{\textbf{\#}} & \multicolumn{2}{l|}{\textbf{Malware}}        & \multicolumn{2}{l|}{\textbf{\#}} & \multicolumn{1}{l|}{\textbf{Benignware}}     & \multicolumn{2}{l|}{\textbf{\#}} & \multicolumn{1}{l|}{\textbf{Malware}}                   & \textbf{\#} \\ \hline
\multicolumn{2}{|l|}{Activity}               & \multicolumn{2}{l|}{1,072,155}     & \multicolumn{2}{l|}{Activity}                & \multicolumn{2}{l|}{2,720,802}     & \multicolumn{1}{l|}{VIEW}                    & \multicolumn{2}{l|}{93,533}       & \multicolumn{1}{l|}{VIEW}                               & 135,251      \\ \hline
\multicolumn{2}{|l|}{Service}                & \multicolumn{2}{l|}{238,136}      & \multicolumn{2}{l|}{Service}                 & \multicolumn{2}{l|}{387,539}      & \multicolumn{1}{l|}{MAIN}                    & \multicolumn{2}{l|}{69,091}       & \multicolumn{1}{l|}{MAIN}                               & 57,902       \\ \hline
\multicolumn{2}{|l|}{Broadcast Receiver}     & \multicolumn{2}{l|}{172,222}      & \multicolumn{2}{l|}{Broadcast Receiver}      & \multicolumn{2}{l|}{212,052}      & \multicolumn{1}{l|}{MESSAGING\_EVENT}        & \multicolumn{2}{l|}{22,181}       & \multicolumn{1}{l|}{CONNECTIVITY\_CHANGE}               & 36,872       \\ \hline
\multicolumn{2}{|l|}{Conent Provider}        & \multicolumn{2}{l|}{96,457}       & \multicolumn{2}{l|}{Conent Provider}         & \multicolumn{2}{l|}{149,763}      & \multicolumn{1}{l|}{BOOT\_COMPLETED}         & \multicolumn{2}{l|}{20,984}       & \multicolumn{1}{l|}{BOOT\_COMPLETED}                    & 28,689       \\ \hline
\multicolumn{2}{|l|}{/}                      & \multicolumn{2}{l|}{/}           & \multicolumn{2}{l|}{/}                       & \multicolumn{2}{l|}{/}           & \multicolumn{1}{l|}{RECEIVE}                 & \multicolumn{2}{l|}{18,204}       & \multicolumn{1}{l|}{USER\_PRESENT}                      & 23,371       \\ \hline
\multicolumn{2}{|l|}{/}                      & \multicolumn{2}{l|}{/}           & \multicolumn{2}{l|}{/}                       & \multicolumn{2}{l|}{/}           & \multicolumn{1}{l|}{CHOOSER}                 & \multicolumn{2}{l|}{17,334}       & \multicolumn{1}{l|}{PACKAGE\_REMOVED}                   & 21,449       \\ \hline
\multicolumn{2}{|l|}{/}                      & \multicolumn{2}{l|}{/}           & \multicolumn{2}{l|}{/}                       & \multicolumn{2}{l|}{/}           & \multicolumn{1}{l|}{INSTANCE\_ID\_EVENT}     & \multicolumn{2}{l|}{13,911}       & \multicolumn{1}{l|}{PACKAGE\_ADDED}                     & 13,972       \\ \hline
\multicolumn{2}{|l|}{/}                      & \multicolumn{2}{l|}{/}           & \multicolumn{2}{l|}{/}                       & \multicolumn{2}{l|}{/}           & \multicolumn{1}{l|}{CONNECTIVITY\_CHANGE}    & \multicolumn{2}{l|}{12,664}       & \multicolumn{1}{l|}{com.taobao.accs.intent.action.RECEIVE}                & 8,856        \\ \hline
\multicolumn{2}{|l|}{/}                      & \multicolumn{2}{l|}{/}           & \multicolumn{2}{l|}{/}                       & \multicolumn{2}{l|}{/}           & \multicolumn{1}{l|}{REGISTER}                & \multicolumn{2}{l|}{11,680}       & \multicolumn{1}{l|}{ACTION\_POWER\_CONNECTED}           & 8,788        \\ \hline
\multicolumn{2}{|l|}{/}                      & \multicolumn{2}{l|}{/}           & \multicolumn{2}{l|}{/}                       & \multicolumn{2}{l|}{/}           & \multicolumn{1}{l|}{INSTALL\_REFERRER}       & \multicolumn{2}{l|}{11,300}       & \multicolumn{1}{l|}{ACTION\_POWER\_DISCONNECTED}        & 8,738        \\ \hline
\end{tabular}}
\end{table*}

To address the problem above, combining the semantics of different types of program flows offers a promising direction for capturing malicious behaviors more effectively.
Specifically, each flow type provides a distinct view for characterizing app behaviors.
For example, a data-flow path can explicitly trace how sensitive data moves from a source to a sink~\cite{DBLP:conf/pldi/ArztRFBBKTOM14}, providing crucial insights into potential data leaks.
Each individual view is capable of profiling particular characteristics of malware but is limited by its scope of assessment.
Therefore, properly integrating the semantics of multiple flow types holds promise for leveraging their complementary strengths to enhance the understanding of app behaviors and uncovering hidden maliciousness within app code.

Existing detection methods with multi-view fusion techniques struggle to model flow-related information accurately, which in turn weakens the expressiveness of the fused representations.
Many methods do not explicitly model the relations between the flow-based features across different views.
Instead, they utilize only partial aspects of flow-related features indirectly, \eg, sensitive API usage, function calls~\cite{kim2018multimodal,qiu2022cyber,gao2023obfuscation,gong2024sensitive}. 
Given the continuity and interdependence of flow characteristics within the views, the indirect modeling hinders the extraction of meaningful semantic associations.
Furthermore, the methods that explicitly represent flow-related information as pre-defined graph structures often overlook the heterogeneity inherent in various flows~\cite{narayanan2018apk2vec,narayanan2018multi}, which potentially leads to the loss of critical behavioral semantics.
For instance, source APIs, sink APIs, and ICC links contribute differently to the semantics of data flows and should therefore be weighted accordingly in maliciousness analysis.

Directly fusing the semantics of different types of flows using existing heterogeneous graph-based methods~\cite{hou2017hindroid,fan2018gotcha,hou2021disentangled,hei2021hawk,shen2024ghgdroid} is technically non-trivial.
Specifically, the semantics of a program flow depend on its context, \eg, the partial order among its constituent entities.
However, the methods above are typically context-unaware, making it difficult to distinguish entities of the same type that appear in different relations, which will be illustrated in \Cref{sec:aware}.
As a result, they fail to capture the mutual constraints and dependencies across the flows, and consequently allow unrelated flows from different apps to be mistakenly combined, leading to semantic confusion and compromising the reliability of the learned representations.

We propose and implement \textit{\oursystem}, a novel technique that achieves context-aware fusion of heterogeneous flow semantics for Android malware detection.
Our primary goal is to leverage complementary strengths of the three types of flow-related semantic information above for profiling app precisely.
Specifically, we explicitly model the relations among entities from various flows using a heterogeneous information network (HIN)~\cite{sun2012mining}.
To incorporate high-level semantics for establishing inter-app relatedness, we construct a meta-path group for each view.
Each group consists of content-oriented and action-oriented meta-paths designed to characterize components, structures, or usage patterns of flows.
To address the computational and storage costs of mining HIN and integrate context awareness, we propose a new HIN embedding approach named \textit{flow2vec}, which learns low-dimensional representations for HIN nodes while accurately preserving both the structural and semantic properties of flows.
Unlike existing techniques~\cite{grover2016node2vec,dong2017metapath2vec,wang2019heterogeneous}, our approach effectively distinguishes the semantics of HIN entities based on contextual constraints across different flows as needed and jointly samples the HIN along multiple meta-paths to enhance app representation learning.
We finally develop a channel-attention-based model to fuse semantic embeddings from the views, weighting their contributions for precise detection.

\begin{figure*}[t] 
\center{\includegraphics[width=0.97\textwidth]  {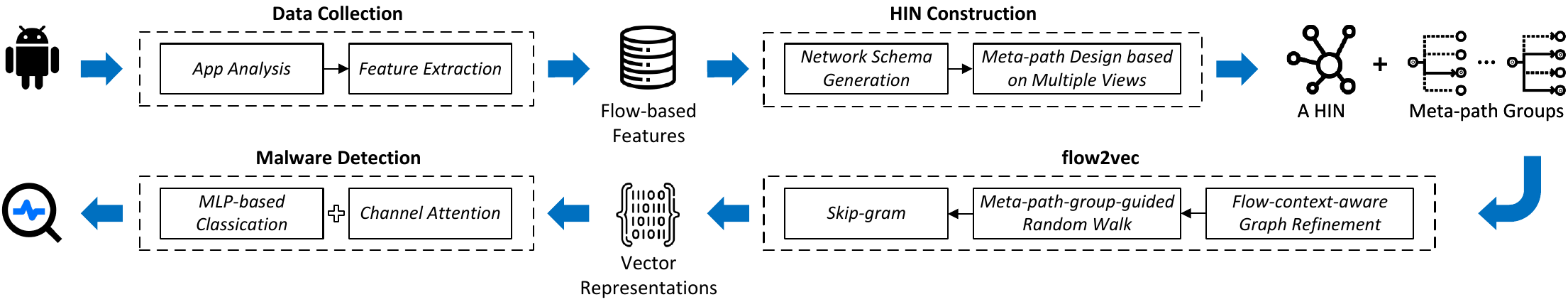}}
\caption{Overall architecture of \oursystem.}
\label{architecture}
\end{figure*}

Our main contributions are summarized as follows:
\begin{itemize}
    \item We propose and implement \oursystem, a novel technique that achieves context-aware fusion of heterogeneous flow semantics for Android malware detection. To the best of our knowledge, it is the first study to comprehensively aggregate the strengths of diverse program-flow-related information for assessing maliciousness of apps.
    \item We design a new context-aware HIN embedding approach named \textit{flow2vec}, which distinguishes the semantics of HIN entities based on contextual constraints across different flows and learns accurate app representations through the joint use of multiple meta-paths.
    \item Our comprehensive evaluations on over 31,000 real-world apps and more than 20 millions program flow instances demonstrate that \oursystem outperforms representative baselines and validate the effectiveness of
    \textit{flow2vec} in representing flow-related semantics.
\end{itemize}

The rest of the paper is organized as follows: Section II introduces our motivation.
Section III explains \oursystem's architecture and the detailed methodology of \oursystem. 
Section IV presents experimental results on \oursystem. 
Section V discusses the limitations and future work. 
Section VI shows the related work, and we conclude in Section VII.

\section{Motivation}\label{sec:motivation}
\iffalse
\begin{table}[t]
\caption{Top 10 actions of Intents in benignware and malware}\label{tab:ACTION}
\resizebox{\linewidth}{!}{
\begin{tabular}{|l|c|l|c|}
\hline
\textbf{Action in Benignware}                    & \textbf{\#}    & \textbf{Action in Malware}                                    & \textbf{\#}     \\ \hline
android.intent.action.VIEW              & 93533 & android.intent.action.VIEW                           & 135251 \\ \hline
android.intent.action.MAIN              & 69091 & android.intent.action.MAIN                           & 57902  \\ \hline
com.google.firebase.MESSAGING\_EVENT    & 22181 & android.net.conn.CONNECTIVITY\_CHANGE                & 36872  \\ \hline
android.intent.action.BOOT\_COMPLETED   & 20984 & android.intent.action.BOOT\_COMPLETED                & 28689  \\ \hline
com.google.android.c2dm.intent.RECEIVE  & 18204 & android.intent.action.USER\_PRESENT                  & 23371  \\ \hline
android.intent.action.CHOOSER           & 17334 & android.intent.action.PACKAGE\_REMOVED               & 21449  \\ \hline
com.google.firebase.INSTANCE\_ID\_EVENT & 13911 & android.intent.action.PACKAGE\_ADDED                 & 13972  \\ \hline
android.net.conn.CONNECTIVITY\_CHANGE   & 12664 & com.sina.weibo.sdk.action.ACTION\_SDK\_REQ\_ACTIVITY & 10172  \\ \hline
com.google.android.c2dm.intent.REGISTER & 11680 & com.taobao.accs.intent.action.RECEIVE                & 8856   \\ \hline
com.android.vending.INSTALL\_REFERRER   & 11300 & android.intent.action.ACTION\_POWER\_CONNECTED       & 8788   \\ \hline
\end{tabular}}
\end{table}
\fi

To motivate our work, we conduct detailed statistical analysis of heterogeneous flows, including control flows, data flows, and ICCs on large-scale apps, and then explain why the three types of flows contribute to effective malware detection.
Specifically, we collect 31,301 real-world samples, including 16,667 benignware samples and 14,634 malware
samples, spanning multiple years from AndroZoo~\cite{allix2016androzoo}.
We then use publicly available tools~\cite{meng2019appscalpel,DBLP:conf/pldi/ArztRFBBKTOM14,DBLP:conf/icse/0029BBKTARBOM15,lam2011soot} to extract triggering conditions, the guarded APIs, source APIs, sink APIs, components, and actions of Intents corresponding to the three types of flows respectively.
The details of our dataset are depicted in \Cref{chap:setup}.

\Cref{stats} lists top-10 key elements extracted from 2018-2022 samples, where common prefixes (\eg, \textit{java.lang}, \textit{android.content}) are removed for brevity.
Based on that, we make the following observations:

(a) There exist clear distinctions between benignware and malware in triggering conditions and guarded APIs.
Malware show higher frequencies of control logic associated with sensitive contexts, \eg, \textit{NETWORK\_INFORMATION} and \textit{UNIQUE\_IDENTIFIER}, compared to benignware. 
Moreover, APIs about file and network operations, \eg, \textit{FileOutputStream.write} and \textit{URLConnection.setRequestProperty}, are more frequently guarded in malware. 
In contrast, benignware tend to guard APIs of general functionalities (\eg, \textit{HashMap.put}, \textit{Activity.getIntent}).
The observations indicate that the combination of triggering conditions and guarded APIs provides discriminative features for malware detection.

(b) The differences emerge in the usage of source APIs and sink APIs between benignware and malware.
While benignware predominantly access standard reflection and collection sources, \eg, \textit{Class.getName} and \textit{HashMap.get}, malware exhibit increased reliance on sensitive sources like \textit{GregorianCalendar.get}, and \textit{SQLiteDatabase.query}. 
In the aspect of sink APIs, benign usage is dominated by common data manipulation APIs, \eg, \textit{HashMap.put} and \textit{String.substring}, whereas malware frequently invoke sensitive sinks, \eg, \textit{JSONObject.put}, \textit{FileOutputStream.write}, and \textit{Camera.setPreviewSize}. 
This suggests that sources and sinks provide meaningful features for malware detection.

(c) The analysis of ICCs reveals divergence in how benignware and malware utilize components and actions of Intents.
Malware invoke more components, particularly Activity and Service, which suggest a higher reliance on ICCs to coordinate background or stealthy behaviors.
Moreover, malware frequently register for system-level actions, \eg, \textit{CONNECTIVITY\_CHANGE}, and \textit{PACKAGE\_REMOVED}, which are less used by benignware. 
The patterns indicate that the ICC-based features provide discriminative clues for malware detection.

(d) The three types of flows exhibit clear semantic relations and complementarity: \textcolor{black}{
\ding{172} Based on the real-world cases, the complementary nature of these flows is supported by semantic intersections across different feature spaces.}
On the one hand, there exists an intersection between guarded APIs and source/sink APIs.
For example, \textit{FileOutputStream.write} appears in both guarded APIs and sink APIs of malware.
\textit{Class.getName} occurs in both guarded APIs and source APIs of benignware.
On the other hand, the components of different types of flows are also explanatorily related.
For instance, \textit{CONNECTIVITY\_CHANGE} in actions of Intents of malware can be seen as an explanatory indicator for the use of a sink API named \textit{URL.openConnection}.
\textcolor{black}{
\ding{173} Theoretically, the three types of flows capture orthogonal yet complementary aspects of app behaviors.
Control flows reflect the structural logic of execution, data flows reveal how sensitive information propagates, and ICCs capture interactions across components, which are frequently exploited in Android malware~\cite{DBLP:conf/icse/AvdiienkoKGZARB15,samhi2022difuzer,xu2016iccdetector}.
While each type of flow provides valuable semantics individually, their integration enables the discovery of complex malicious patterns that span multiple behavioral dimensions, which may not be fully captured by any single feature type alone.
For example, different types of flows can enrich the semantics of sensitive APIs by providing complementary context (\ie, execution conditions, execution purposes, and data transmission modes) to help characterize how the APIs are used.
This justifies the integration of heterogeneous flows as a principled and effective choice for malware behavior analysis.
}

% Based on the observations, it is evident that each selected view offers unique advantages in identifying Android malware. 
% Therefore, effectively leverage the complementary relations among different views can help capture the high-level behavioral semantics of apps, thereby enabling precise Android malware detection.
%!TEX root = bare_jrnl.tex

\section{Methodology}
% This section details the implementation of \oursystem. It begins with an overview of its architecture, followed by the process of feature extraction from different views. Next, it describes the construction of HIN, including the network schema and meta-paths derived from the features. The representation of HIN nodes for Android malware detection is then explained. Finally, it presents the DNN-based classifier used to identify malware.

\subsection{Architecture}

\Cref{architecture} depicts the overall architecture of \oursystem, consisting of four main components as follows: 
% processes: data modeling and malware detection.
% The former builds a HIN based on different types of behavioral analysis results from apps with the specified analyzers.
% The latter generates vectorial representations for the apps based on the HIN and then trains \textcolor{red}{an attention-based DNN classifier} by feeding the embedding results to detect malware. 
% The above processes are divided into four components:

\textbf{Data Collection.} The component obtains flow-based features by off-the-shelf static analysis tools. 
Specifically, \oursystem employs IccTA~\cite{DBLP:conf/icse/0029BBKTARBOM15} to collect explicit and implicit ICC links.
Additionally, it uses FlowDroid~\cite{DBLP:conf/pldi/ArztRFBBKTOM14} to gather intra- and inter-component sensitive data-flow paths.
Furthermore, it gets conditional statements of the guarded sensitive APIs by the Soot framework~\cite{lam2011soot,meng2019appscalpel}.
Based on the raw data above, \oursystem automatically extracts meaningful features (\eg, trigger conditions, sensitive APIs), and then analyzes various relations (\eg, \textit{condition-trigger-API}) among different types of entities (\eg, condition, API).

\textbf{HIN Construction.} The component constructs a HIN using the flow-based features extracted by the previous component.
Specifically, a network schema is generated to model the relations among various entities of heterogeneous flows.
Based on the schema, the HIN is constructed.
After that, three groups of meta-paths are designed from the HIN to capture the relatedness over apps from different views of flows.
To simplify our design, each group of meta-paths include one content-oriented meta-path and one action-oriented meta-path.
    
\textbf{flow2vec.} The component utilizes a context-aware HIN embedding approach to generate low-dimensional representations of nodes in the HIN, while accurately preserving both semantics and structural correlations between different types of nodes.
To ensure the semantic correctness of flow-related embeddings on the HIN, a flow-context-aware graph refinement method is proposed to separate nodes based on their associated contexts.
Next, a random walk strategy guided by predefined meta-path groups is designed to map the word-context concept in a text corpus into the HIN.
This strategy guarantees the semantic completeness of flow-related information during the sampling process on the HIN.
Skip-gram is finally used to learn node representations for the HIN.
    
\textbf{Malware Detection.} The component trains a channel-attention-based deep neural network (DNN) model using the vectors learned by \textit{flow2vec} to determine whether a given app is malicious.
Specifically, a channel attention mechanism is used to dynamically adjust the importance of the vectors from different views, thereby enhancing the effectiveness of their fusion.
    % fectiveness of the downstream detection task.
Finally, a multi-layer perceptron(MLP)-based classifier is trained to produce a predicted label for the input app.

\subsection{Data Collection} \label{featureExtraction}
Based on the domain knowledge, we leverage existing static analysis tools to extract features from three views centered on control flows, data flows, ICCs.
The flow-based features contain various entities of apps (\eg, APIs, conditions, actions) and rich semantic relations among them, all of which are crucial to build the HIN in the following. 
% Note that we adopt the typical views to validate the effectiveness of \oursystem empirically, while our scheme can be extended to work based on the detection results from the other views~\cite{narayanan2018apk2vec,arp2014drebin,enck2014taintdroid}.

\subsubsection{Feature Extraction from Control-flow View} Using control structures for sensitive operations is one of the common tricks by which malware can be camouflaged as benignware~\cite{pan2017dark,samhi2022difuzer}.
To capture this kind of malicious intentions, we obtain trigger conditions, as well as the APIs guarded by the conditions within apps as follows.

We leverage the Soot framework to obtain the conditions that control the executions of sensitive APIs.
Specifically, app code is first transformed into Jimple, the internal representation of Soot~\cite{lam2011soot}. 
Sensitive APIs are then positioned in the Jimple code based on their API signatures.
We extract the trigger conditions and their triggering semantics based on AppScalpel's strategy~\cite{meng2019appscalpel} implemented by Soot APIs as follows.

We position the target conditional statement based on a rule that the target statement is the predominator of an invocation statement of the guarded sensitive API, but the invocation statement is not its postdominator, and meanwhile, the join point of the conditional statement's branches is the postdominator of the invocation statement.
We next extract the semantics of the obtained conditions by code instrumentation.
Specifically, the functionality of a source API that has information flow to a conditional statement can be regarded as the semantic of that statement.
However, conditional statements cannot be considered as sinks when using state-of-the-art static taint analyzers~\cite{DBLP:conf/pldi/ArztRFBBKTOM14,wei2018amandroid}.
To address this limitation, \textcolor{black}{we insert a call to a dummy method \textit{ifStmt()} before each conditional statement.
The call includes the variables involved in the condition as its parameters.
The \textit{ifStmt()} is static and declared in a dummy class \textit{IfClass} that contains all the instrumented methods related to conditions.
As depicted in Listing \ref{codeinst} for an app\footnote{\texttt{MD5: dc9b09466a0024f22989f71d9632a0d6}}, we add code at Line 8 to wrap the parameters (\ie, \textit{LocalLocationProvider} and \textit{localObject2}) used in the conditional statement at Line 9.}

\begin{lstlisting}[caption={\textcolor{black}{A simplified code snippet of a real-world malware sample named \textit{com.Gamezoor.Grand.CityRacing.game} for the instrumentation performed by MalFlows (a Line with "+" represents the instrumented code).}}, label={codeinst}]
1  public final void b() {
2     LocationProvider localLocationProvider;
3     do {
4        localLocationProvider = localLocationManager.getProvider((String)((Iterator)localObject2).next());
5     } while(localLocationProvider.getAccuracy() != 2);
6     ...
7     // A dummy method call for a conditional statement
8 +   IfClass.ifStmt(@\textbf{localLocationProvider, localObject2}@);
9     if((@\textbf{localLocationProvider}@ == null) || (localLocationManager.getProvider((String)@\textbf{localObject2}@).getAccuracy() != 1)) {
10        a(localLocationManager.getLastKnownLocation(...);
11    }
12 }
\end{lstlisting}

\textcolor{black}{Upon completion of the instrumentation process, we dynamically designate each newly inserted method call as a sink to facilitate downstream analysis.}
Based on that, we can indirectly get the data flows from source APIs and \textit{ifStmt()}.
We finally leverage the categories proposed by SuSi~\cite{DBLP:conf/ndss/RasthoferAB14} to summarize the semantics of the condition-related sources.
For example, the semantics of \textit{getDeviceId()} and \textit{getSubscriberId()} are summarized as \textit{UNIQUE\_IDENTIFIER}.

% Furthermore, we also record the apps using the APIs.
% Note that we leverage the context factors summarized by AppContext~\cite{yang2015appcontext} to depict the trigger conditions.

We build two matrixes to describe the relations above: 
\begin{itemize}
    \item \textbf{R$_1$:} The \textit{app-include-condition} matrix \textbf{C} where each element $c_{i,j} \in \{0, 1\}$ means if app$_i$ includes condition$_j$.
    \item \textbf{R$_2$:} The \textit{condition-trigger-API} matrix \textbf{T} where each element $t_{i,j} \in \{0, 1\}$ means if condition$_i$ controls the invocation of sensitive API$_j$.
\end{itemize}

\subsubsection{Feature Extraction from Data-flow View} \label{data_flow_view} Usage of sensitive data within apps is one of the most important clues to identify hidden maliciousness~\cite{DBLP:conf/pldi/ArztRFBBKTOM14,DBLP:conf/icse/AvdiienkoKGZARB15}.
We run FlowDroid~\cite{DBLP:conf/pldi/ArztRFBBKTOM14} to collect intra- and inter-component data-flow paths for each app.
We then get a pair of APIs (\ie, a source API and a sink API) from each path and record the apps using the APIs.

We build two matrixes to represent the obtained relations:
\begin{itemize}
    \item \textbf{R$_3$:} The \textit{app-use-API} matrix \textbf{U} where each element $u_{i,j} \in \{0, 1\}$ means if app$_i$ uses API$_j$.
    \item \textbf{R$_4$:} The \textit{API-flow-API} matrix \textbf{F} where each element $f_{i,j} \in \{0, 1\}$ means if there exists a data flow from API$_i$ to API$_j$. API$_i$ is a source API and API$_j$ is a sink API.
\end{itemize}

% \begin{table}[tb]
% \centering
% \caption{Features extracted from different views.}
% \resizebox{\linewidth}{!}{
% \begin{tabular}{|l|l|l|l|}
% \hline
% \textbf{View}       & \textbf{Relation}     &  \textbf{Description}   \\ \hline
% \multirow{2}{*}{Data Flow}   & app-use-API  &  An app uses an API  \\ \cline{2-3}
% & API-flowto-API &  A data flow exists from an API to another  \\ \hline
% \multirow{2}{*}{Control Flow}    &  app-include-cond &  An app includes a conditional statment    \\ \cline{2-3}
% & cond-control-API &  A condition controls the invocation of an API \\ \hline
% ICC Link            &          &              \\ \hline
% \multirow{2}{*}{Permission}        & app-declare-perm &  A permission is declared in an app      \\ \cline{2-3}
% &  perm-regulate-API  &  A permission regulates the execution of an API \\ \hline
% \end{tabular}
% }
% \label{tab:features}
% \end{table}

% the \textit{app-include-condition} matrix \textbf{I} where each element $i_{i,j} \in \{0, 1\}$ means if app$_i$ includes condition$_j$ (\textit{R3}), and the \textit{condition-trigger-API} matrix \textbf{T} where each element $t_{i,j} \in \{0, 1\}$ means if condition$_i$ controls the execution of API$_j$ (\textit{R4}), and the \textit{app-use-API} matrix \textbf{U} mentioned in \Cref{data_flow_view}.

\subsubsection{Feature Extraction from ICC View} The ICC mechanism can be exploited by malware to launch stealthy attacks, \eg, data leaks, code obfuscation, privilege escalation~\cite{xu2016iccdetector,DBLP:conf/icse/0029BBKTARBOM15}.
To enhance the detectability of \oursystem for ICC-based malicious behaviors, we first run IccTA~\cite{DBLP:conf/icse/0029BBKTARBOM15} to get ICC links from apps and extract actions of Intents and source components from each ICC link.
The action is a core and widely used attribute of an implicit Intent, informing what operation is expected to be executed.
The source component refers to the sender of an explicit or impilict Intent.
The two types of entities are essential to interpret the semantics of ICC links.

We build two matrixes to represent the obtained relations:
\begin{itemize}
    \item \textbf{R$_5$:} The \textit{app-set-action} matrix \textbf{S} where each element $s_{i,j} \in \{0, 1\}$ means if action$_j$ is set in app$_i$.
    \item \textbf{R$_6$:} The \textit{app-declare-component} matrix \textbf{D} where each element $d_{i,j} \in \{0, 1\}$ means if app$_i$ declares component$_j$ in AndroidManifest.xml.
    \item \textbf{R$_7$:} The \textit{component-initiate-action} matrix \textbf{N} where each element $n_{i,j} \in \{0, 1\}$ means if component$_i$ initiates action$_j$ in an Intent.
\end{itemize}

\iffalse
\subsubsection{Extraction from the Permission View} The weakness of Android users in understanding and using permissions gives malicious developers an opportunity to harm users by abusing permissions~\cite{arora2019permpair}.
We thus use PScout~\cite{au2012pscout} to gather API-permission mappings from apps.

To depict the aforementioned relations, we build three matrixes: the \textit{app-declare-permission} matrix \textbf{D} where each element $d_{i,j} \in \{0, 1\}$ means if permission$_j$ is declared in app$_i$ (\textit{R7}), and the \textit{permission-regulate-API} matrix \textbf{R} where each element $r_{i,j} \in \{0, 1\}$ means if permission$_i$ regulates API$_j$, and the \textit{app-use-API} matrix \textbf{U} mentioned in \Cref{data_flow_view}.
\fi

\begin{figure}[tb] 
\center{\includegraphics[width=0.76\linewidth]  {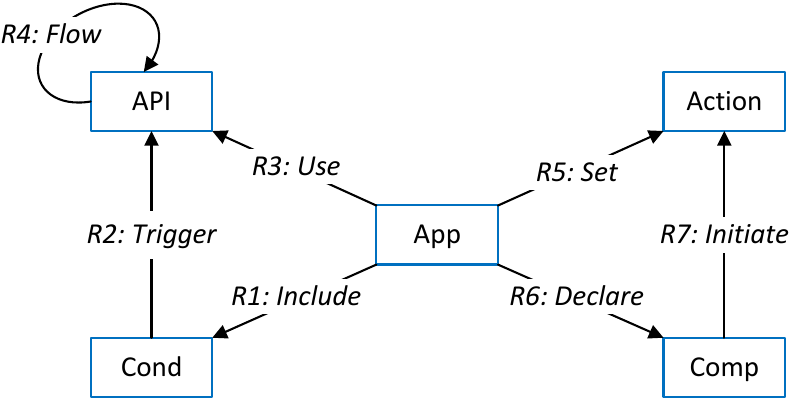}}
\caption{Network schema for the HIN in \oursystem, where blue solid rectangles represent entity types, and black solid arrows indicate their relations.}
\label{HINschema}
\end{figure}

\subsection{HIN Construction} \label{HINConstruction}
To depict the rich relations among the extracted Android entities, it is crucial to model them in an appropriate manner so that different relations can be better and easier analyzed and handled.
Therefore, we employ a HIN~\cite{sun2012mining} that is capable to incorporate different types of features (\ie, entities and relations) extracted above.
The HIN facilitates uncovering underlying intentions within Android apps by providing not only the network structure of the data associations but a high-level abstraction of categorical association.
We first exhibit the definitions of the HIN and its network schema as follows.  
\begin{myDef}
    A heterogeneous information network~\cite{sun2012mining} is defined as a graph G = (V, E) with an entity type mapping $\phi$: V $\rightarrow$ T and a relation type mapping $\psi$: E $\rightarrow$ R, where V denotes the entity set and T denotes the entity type set, and E denotes the relation set and R denotes the relation type set, and the number of entity types $\lvert$T$\rvert$ $\textgreater$ 1 or the number of relation types $\lvert$R$\rvert$ $\textgreater$ 1. The network schema for G, denoted as S$_{G}$ = (T, R), is a graph with nodes as entity types from T and edges as relation types from R.
\end{myDef}

\subsubsection{Network Schema Generation}
We have 5 entity types (\eg, APIs, conditions) and 7 types of relations among them (\ie, R$_1$-R$_7$ in \Cref{featureExtraction}).
The network schema for the HIN of our work is shown in~\Cref{HINschema}, which enables apps to be comprehensively represented by simultaneously incorporating information from multiple views.

\begin{figure}[tb] 
\center{\includegraphics[width=1
 \linewidth]  {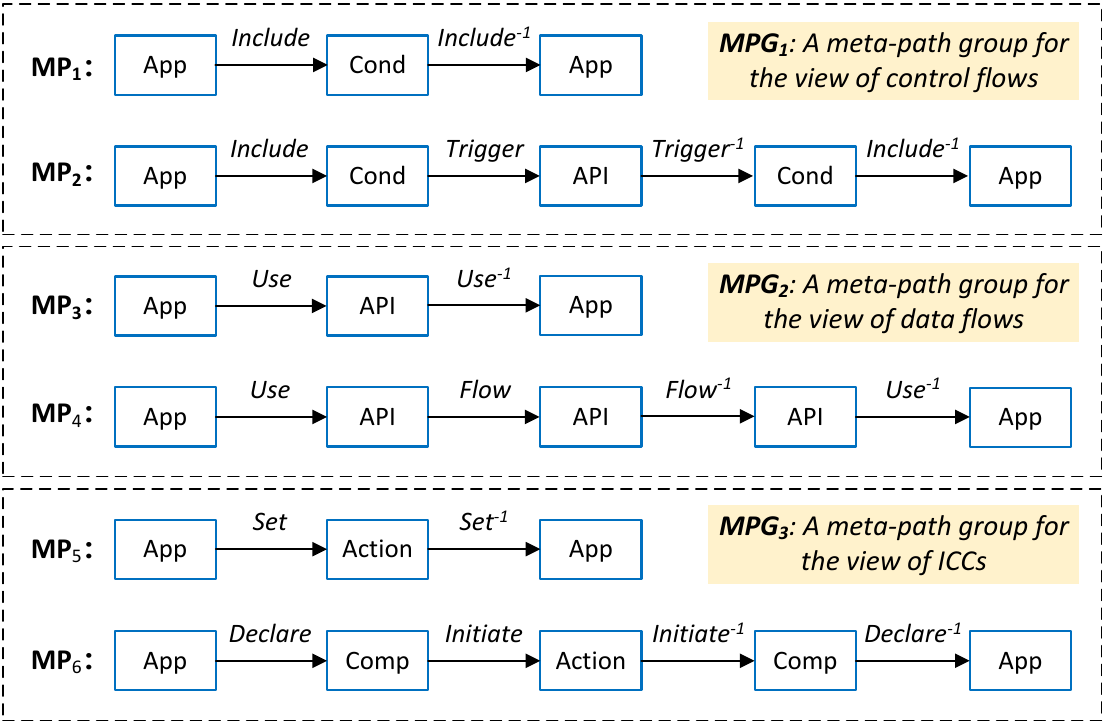}}
\caption{Meta-paths built for flow-centric malware detection, where some entities (\eg, \textit{Cond}) is short for the contents in \Cref{HINschema} (\eg, \textit{Condition}).
The definitions of nodes and edges are the same as those in \Cref{HINschema}.}
\label{metapaths}
\end{figure}

\subsubsection{Meta-path Design}\label{chap:metapath}
The different types of entities and relations motivate us to use a machine-readable representation for enriching the semantics of similarities among apps.
The similarities are our basis to measure the maliciousness of apps.
To handle this, the meta-path is designed to formulate the semantics of higher-order relations among entities in HIN.
Its definition is shown as follows.
% Here, we follow this concept, which is defined as follows, and extend it to our work in Android malware detection.

\begin{myDef}
    A meta-path MP is a path defined on the graph of network schema S$_{G}$ = (T, R), and is denoted in the form of T$_{1}$ $\stackrel{R_{1}}{\longrightarrow}$ T$_{2}$ $\stackrel{R_{2}}{\longrightarrow}$ T$_{3}$ $\stackrel{R_{3}}{\longrightarrow}$ ... $\stackrel{R_{M}}{\longrightarrow}$           
                              T$_{M+1}$, which defines a composite relation R = R$_{1}$ $\circ$ R$_{2}$ $\circ$ R$_{3}$ $\circ$ ... $\circ$ R$_{M}$ between types T$_{1}$ and T$_{M+1}$, where $\circ$ denotes relation composition operator, and M is the length of MP.
\end{myDef}

Given the network schema with different types of entities and relations, we enumerate many meta-paths.
Based on the domain knowledge from human experts, we design 6 meaningful meta-paths shown in \Cref{metapaths} for characterizing relatedness over apps.
To aggregate different types of semantic information from our HIN, we then group the meta-paths based into 3 distinct but related views, \ie, control flows, data flows, and ICCs.
Each group comprises one content-oriented meta-path (\eg, MP$_1$) and one action-oriented meta-path (\eg, MP$_2$), which work together to capture various relations within a view.
Note that additional meta-paths can be included in each group, but we select the two most representative ones to demonstrate the effectiveness of \oursystem.

For example, the relatedness over apps by leveraging the control-flow features in a meta-path MP$_1$ is 
\textit{App} $\xlongrightarrow{\text{Include}}$ \textit{Condition} $\xlongrightarrow{\text{Include}^{\text{-1}}}$ \textit{App}, which means that two apps can be connected as they use the trigger conditions with the identical semantics.
Another meta-path MP$_2$ in the same group, \ie, 
\textit{App} $\xlongrightarrow{\text{Include}}$ \textit{Condition} $\xlongrightarrow{\text{Trigger}}$ \textit{API} $\xlongrightarrow{\text{Trigger}^{\text{-1}}}$ \textit{Condition} $\xlongrightarrow{\text{Include}^{\text{-1}}}$ \textit{App}, denotes that two apps are related as they use the trigger conditions controlling the invocations of the same API.
The former depicts the contents of apps, but the latter describes the detailed actions of apps.

% Furthermore, meta-paths in the same group exhibit complementarity, as they provide additional information to each other and remove potential errors in semantic inference, which will be explained below.

\subsection{flow2vec} \label{HINRepresentation}
To measure the relatedness over entities of the constructed HIN, it is essential to adopt an effective representation learning method that effectively aggregates both structural and semantic relations about various flows.
\textcolor{black}{
We thus propose a new HIN embedding approach named \textit{flow2vec}, which consists of two main modules: a flow-context-aware graph refinement module, and a meta-path-group-guided random walk module.
To maintain a clear and modular design, we decouple graph refinement from random walk.
This allows invalid relations to be filtered in advance and enables flexible integration of new graph refinement or graph embedding strategies without altering core components.
}
% To this end, we first formalize the problem of HIN representation learning as follows.

% \begin{myDef}
%     HIN representation learning~\cite{dong2017metapath2vec} is that given a HIN G, the task is to learn a function \emph{f}: \emph{V} $\rightarrow \mathcal{R}^{d}$ that maps each node \emph{v} $\in$ \emph{V} to a vector in a d-dimensional space $\mathcal{R}^{d}$, d $\ll$ $\mid$\emph{V}$\mid$ that are capable to preserve the structural and semantic relations among them.
% \end{myDef}

\subsubsection{Flow-context-aware Graph Refinement}\label{sec:aware}
Modeling flow-related information using a standard HIN method~\cite{hou2017hindroid} directly results in the description of incorrect relations.
% renders conventional random-walk-based methods (\eg, metapath2vec~\cite{dong2017metapath2vec}) inapplicable for learning latent representations.
Specifically, flow-related information in the HIN consists of multiple entities, associated not only through the explicitly modeled relations but also by implicit contextual constraints (\eg, the source API and the sink API in a data-flow path are used in the same app).
Overlooking the constraints lead to incorrect extraction of structural and semantic relations of the HIN.

\begin{figure}[tb]
\centering
\center{\includegraphics[width=\linewidth]  {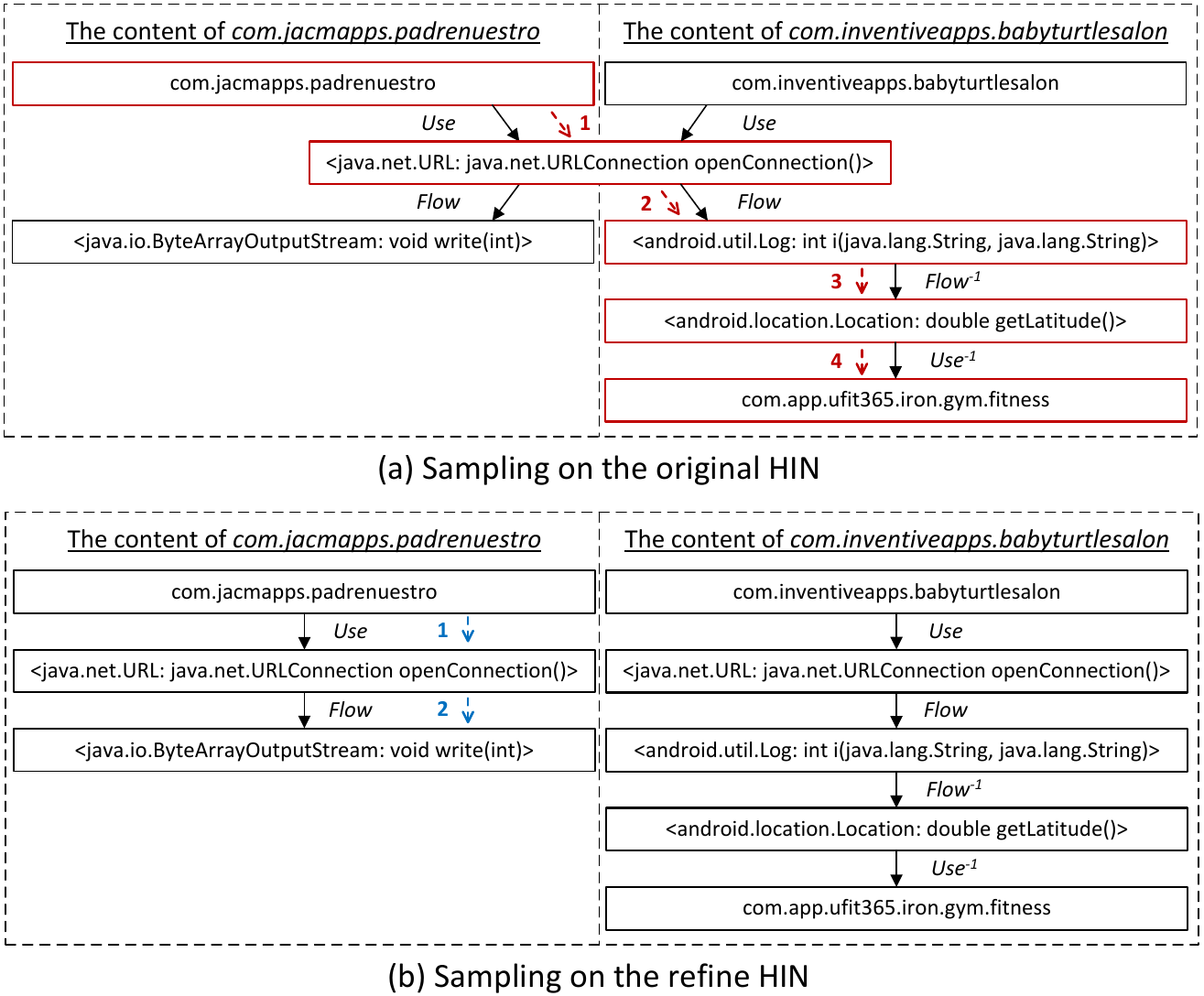}}
\caption{Example for sampling on the original and refined HIN, where solid rectangles represent entities, black solid arrows indicate their relations, and the numbers on the dashed arrows show the order of access.
Incorrect sampling paths are highlighted in red, and the correct paths are highlighted in blue.
}
\label{wrongPath1}
\end{figure}

\begin{figure}[tb] 
\center{\includegraphics[width=\linewidth]  {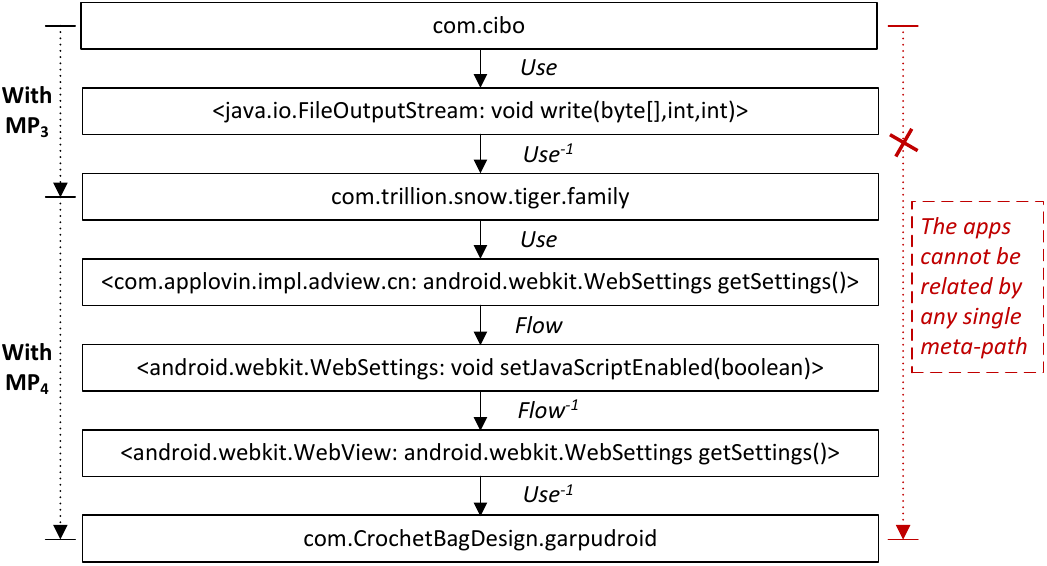}}
\caption{Example of the HIN sampling guided by a meta-path group, where the black solid arrows represent the relations between the entities.}
\label{moti}
\end{figure}

We use a real-world case to demonstrate the importance of modeling the flow-related information under contextual constraints.
As drawn in \Cref{wrongPath1}(a), the red sampling paths on the subgraph of the HIN violate a contextual constraint. 
Specifically, the app named \textit{com.jacmapps.padrenuestro}\footnote{\texttt{MD5: 0bffa8e7b29306f8bbec28dba6bb7150}} and the app named \textit{com.app.ufit365.iron.gym.fitness}\footnote{\texttt{MD5: cbe89039caf326cff16070d50c0c0da8}} appear to be related by MP$_3$ under the data-flow view.
However, the sink API named \textit{$\langle$android.util.Log: int i(java.lang.String, java.lang.String)$\rangle$} is not used in the former app.
Therefore, the data-flow-based relation does not exits in practice.
The former app and the sink API are incorrectly associated over the HIN, resulting in inaccurate representations for the two apps.
In comparison, as shown in \Cref{wrongPath1}(b), by representing the API named \textit{$\langle$java.net.URL: java.net.URLConnection openConnection()$\rangle$} as two separate instances corresponding to different apps, the graph refinement enables the HIN to capture context-specific semantics and prevents unrelated apps from being incorrectly linked.
% the API named \textit{$\langle$java.net.URL: java.net.URLConnection openConnection()$\rangle$} is split into two APIs belonging to different apps, making the two apps are not related on the HIN.
All the paths (\eg, the blue one) sampled from the refined HIN exist in practice.

% \begin{figure}[tb] 
% \center{\includegraphics[width=0.82
%  \linewidth]  {../graphs/case.pdf}}
% \caption{Examples for the original and refined HIN, where solid rectangles
% represent entities, and black solid arrows indicate their relations. The numbers on the dashed arrows indicate the order of access.
% Incorrect paths are highlighted in red, and correct paths are highlighted in blue for clarity.}
% \label{wrongcase}
% \end{figure}

% As shown in \Cref{wrongcase}(a), the red paths on the subgraph of the HIN violate a context constraint.
% Specifically, Condition$_1$ is included in App$_2$ and Condition$_1$ controls the invocation of API$_1$, but API$_1$ is not actually used in App$_2$.
% Therefore, this control-flow-based relation highlighted in red does not exist in practice.
% Such semantic incorrectness leads to inaccurate vector representations, which affect downstream detection results.
% In comparison, as depicted in \Cref{wrongcase}(b), Condition$_1$ is split into Condition$_3$ and Condition$_4$, where the former is included in App$_1$ and the latter is included in App$_2$.
% In the refined HIN, all the paths (\eg, the blue one) exist in practice.

As depicted in \Cref{alg:graphrefine}, we leverage joint semantics of meta-paths in each group to identify and correct the inaccurate parts in a HIN on demand.
We first obtain all nodes with the second entity type along \textit{mp\_l} (\ie, anchor nodes) from a HIN \textit{H} (Line 3).
We then find predecessors \textit{P} and successors \textit{S} of each anchor node (Lines 5-6).
Next, we clone each anchor node as \textit{nc} and rebuild its connection relations as follows.
We connect a predecessor \textit{p} to \textit{nc} and \textit{nc} to a successor \textit{s} when \textit{p} and \textit{s} can be connected via any meta-path in \textit{S\_mp\_s} (Lines 9-10).
Otherwise, we connect \textit{p} to \textit{nc} only once (Lines 11-12).
In other word, we use content-oriented relations of the meta-paths in \textit{S\_mp\_s} to constrain action-oriented relations of \textit{mp\_l} in the HIN construction.
Afterwards, we remove the anchor node to avoid information redundancy in \textit{H} (Line 15).
Finally, a refined HIN is generated.

\iffalse
The neural-network-based methods~\cite{wang2019heterogeneous,hu2020heterogeneous} are not suitable to explicitly control meta-paths in the HIN.
On the one hand, these methods cannot directly learn precise representations of context-rich flows.
On the other hand, it would be costly to adapt our work by modifying their models.
% Furthermore, the interpretability of the methods is limited, so the contributions of different meta-paths are hard to be evaluated.
\fi

\begin{algorithm}[t]
\caption{Implementation of the graph refinement}\label{alg:graphrefine}
\label{alg:example}
\begin{algorithmic}[1] % [1] means line numbers will be shown
    \STATE \textbf{Input:} A HIN \textit{H}, a meta-path \textit{mp\_l} of length 4, a set \textit{S\_mp\_s} consisting of meta-paths of length 2
    \STATE \textbf{Output:} A refined HIN \textit{H}  
                \STATE \textit{N} $\gets$ getAnchorNodes(\textit{H}, \textit{mp\_l}) 
                \FORALL{\textit{n} in \textit{N}}
                    \STATE \textit{P} $\gets$ getPredecessor(\textit{n}, \textit{H}, \textit{mp\_l});
                    \STATE \textit{S} $\gets$ getSuccessor(\textit{n}, \textit{H}, \textit{mp\_l})   
                    % obtain the predecessor nodes \textit{P} and successor nodes \textit{S} of \textit{n} in \textit{H} along \textit{mp\_long}
                    \FORALL{\textit{p} in \textit{P} and \textit{s} in \textit{S}}
                        % \FOR{}
                            \STATE
                            \textit{nc} $\gets$ clone(\textit{n})
                            \IF{isConnected(\textit{p}, \textit{s}, \textit{S\_mp\_s})}
                                \STATE setPredecessor(\textit{nc}, \textit{p},  \textit{H}); setSuccessor(\textit{nc}, \textit{s}, \textit{H})       
                            \ELSIF{haveNotConnected(\textit{p}, \textit{H})}
                                \STATE setPredecessor(\textit{nc}, \textit{p},  \textit{H})
                            \ENDIF
                        % \ENDFOR
                        % \STATE clone \textit{n} as \textit{n'} while preserving predecessor and successor relations of \textit{n}, and then add \textit{n'} in \textit{H} 
                        % \STATE remove the edge from \textit{p} to \textit{n} in \textit{H}
                    \ENDFOR 
                    \STATE removeOriginalNode(\textit{n}, \textit{H})
                \ENDFOR
    \RETURN \textit{H}
\end{algorithmic}
\end{algorithm}

\iffalse
faced with the , traditional representation learning methods that mainly factorize the matrix of a HIN to generate latent-dimension features for nodes~\cite{hoff2002latent,sun2011pathsim,zhao2017meta} may incur the expensive cost when decomposing a large-scale matrix, and then also suffer from the statistical performance drawback~\cite{grover2016node2vec}.

One the other hand, it may be infeasible to directly apply the HIN embedding methods, \eg, metapath2vec~\cite{dong2017metapath2vec}, to learn the latent representations for the HIN in our work.
Specifically, Android malware can disguise themselves with different methods, so it is helpful to identify malicious app behaviors by combining the information from multiple views. 
As shown in \textcolor{red}{X}, the apps in the HIN can be associated by combining different meta-paths.
\fi

% Consequently, the metapath2vec that supports traversing the HIN based on one meta-path fails to capture the compound semantic relation.

\subsubsection{Meta-path-group-guided
Random Walk}\label{chap:rw}
To learn latent representations for the refined HIN, it is inapplicable to directly apply traditional single meta-path-based sampling schemes, \eg, metapath2vec~\cite{dong2017metapath2vec}, HAN~\cite{wang2019heterogeneous}, MAGNN~\cite{fu2020magnn}.
% Specifically, current Android malware can adopt various methods to disguise itself, reducing the distinguishability between benignware and malware.
Specifically, single meta-path-based sampling considers only a limited subset of flow-based relations in the HIN, so it may fail to capture important structural and semantic dependencies.
As depicted in \Cref{moti}, malware samples named \textit{com.cibo}\footnote{\texttt{MD5: 5aa13c2dfe6668a6a170e82c70c1a56c}} and \textit{com.CrochetBagDesign.garpudroid}\footnote{\texttt{MD5: 37218220d384bf08aece74ab38479ce4}} cannot be related by any single meta-path in \Cref{chap:metapath}.
Due to the failure to capture their relations, the two apps may be misclassified in malware detection, 
In comparison, the apps can be connected by the meta-paths in MPG$_2$, where the app named \textit{com.trillion.snow.tiger.family}\footnote{\texttt{MD5: 62925eabca26b8be2f6f7a2f280900b0}} bridges them in the HIN.
By capturing the relations, the apps can be correctly classified.

% Moreover, the neural network based methods generate node embeddings separately for different meta-paths and fuse them by weight, but they do not jointly model multiple meta-paths in an integrated manner, which may cause cross-meta-path information loss and negatively impact detection effectiveness.

To produce precise node embeddings for our work, we extend metapath2vec based on the proposed work~\cite{ye2018icsd} for sampling based on each meta-path group.
This facilitates the extraction of structural and semantic information from both contents and actions cooperatively.
Given a meta-path group \textit{MPG}, we put a random walker to traverse a HIN.
The walker randomly chooses a meta-path MP$_{k}$ from \emph{MPG} at first and the transition probabilities at step $i$ are defined as follows:
\resizebox{\linewidth}{!}{
\renewcommand{\arraystretch}{1.3}
$\begin{array}{l}
p\left(v^{i+1} \mid v_{m}^{i}, \emph{MPG} \right) = \\
\left\{
\begin{array}{lr}
\frac{\mu}{\mid\emph{MPG}\mid} \frac{1}{\mid N_{m+1}\left(v_{m}^{i}\right)\mid} & \text { if }\left(v_{m}^{i}, v^{i+1}\right) \in \emph{E}, \phi\left(v_{m}^{i}\right) = T_{app},\\
& \phi\left(v^{i+1}\right) = T_{t+1} \\
\frac{1}{\left|N_{m+1}\left(v_{m}^{i}\right)\right|} & \text { if }\left(v_{m}^{i}, v^{i+1}\right) \in \emph{E}, \phi\left(v_{m}^{i}\right) \neq T_{app}, \\
& \phi\left(v^{i+1}\right) = T_{t+1}, \left(T_{m}, T_{m+1}\right) \in \text{MP}_{k} \\
0 & \text { otherwise,}
\end{array}
\right.
\end{array}$
}
where $\phi$ is the node type mapping function, $v_{m}^{i}$ denotes the $T_{m}$ type of the entity visited at step $i$, $N_{m+1}\left(v_{m}^{i}\right)$ denotes the $T_{m+1}$ type of neighborhood of the entity $v_{m}^{i}$, $T_{app}$ is the entity type of app, and $\mu$ is the number of meta-paths starting with $T_{app} \rightarrow T_{m+1}$ in the given meta-path group \emph{G}.
Note that MP$_{k}$ will be updated by the meta-path drawn from \emph{G} with $\frac{\mu}{\mid\emph{S}\mid}$ probability when the walker meets the case at the first line of the above formula.
\textcolor{black}{To enhance the efficiency of graph embedding in practice, our algorithm supports parallel processing of the meta-path groups derived from three views.}

When the sampling length is long enough, the walker can traverse the HIN with various combinations of meta-paths within the given meta-path group.
For example, if \emph{MPG} = \{MP$_{1}$, MP$_{2}$\}, the walker may traverse based on the combined meta-paths, \eg, [MP$_{1}$, MP$_{1}$, ...], [MP$_{2}$, MP$_{2}$, ...], [MP$_{1}$, MP$_{2}$, MP$_{1}$, MP$_{2}$, ...].
The sampled paths accurately preserve both structural and semantic relations between different types of nodes in the HIN, and thus facilitate the transformation of HIN structures into sequences.

\subsubsection{Skip-gram}
We use skip-gram~\cite{mikolov2013efficient} to generate vectors for entity nodes in the sampled paths.
\textcolor{black}{Skip-gram is originally designed to learn latent representations of words according to their contexts in sentences.
We map the word-context concept in a text corpus into a HIN via the meta-path-group-guided random walk, where a sentence in the corpus corresponds to a sampled path and a word corresponds to an entity node.}

\begin{figure}[tb] 
\center{\includegraphics[width=0.9\linewidth]  {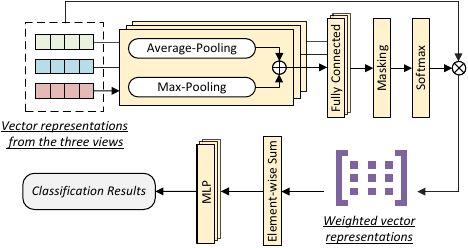}}
\caption{Channel-attention-based deep neural network classifier.}
\label{DNN}
\end{figure}

% \begin{table}[tb]
% \centering
% \caption{The parameters of our model.}
% \resizebox{0.6\linewidth}{!}{
% \begin{tabular}{l|ll}
% \hline
% \textbf{Layer}          & \multicolumn{2}{l}{\textbf{Parameter Settings}} \\ \hline
% Self-Attention &                   &                   \\ \hline
% Input          &                   &                   \\ \hline
% Hidden         &                   &                   \\ \hline
% Hidden         &                   &                   \\ \hline
% Output         &      1 neurons             &       Sigmoid            \\ \hline
% \end{tabular}
% }
% \label{DNNparm}
% \end{table}

\subsection{Channel-attention-based DNN Classifier} \label{DNNClassifier}

To leverage strengths of the semantics of the heterogeneous flows for
profiling app behaviors, we design a channel-attention-based DNN classifier.
As shown in \Cref{DNN}, for an app, we treat each feature vector extracted from a view of flows as a distinct channel and then employ channel attention to adaptively reweight each feature vector for the fusion of the semantics.
This treatment enhances the most informative representations while suppressing less relevant ones, making a more comprehensive and discriminative fused representation for precise malware classification.

\subsubsection{Channel-attention-based Semantic Fusion for Heterogeneous Flows}
We first aggregate semantic information of the feature vectors $V$ $\in \mathbb{R}$$^{c \times d}$ from the three views, where $c$ is the number of channels, and $d$ is the dimension of each input feature.
In our work, $c$ = 3 and $d$ = 128.
\textcolor{black}{Based on the content in \Cref{DNN}, the specified computation process can be represented by the following set of equations:
\begin{align}
z &= V_{\text{avg}} + V_{\text{max}}, \\
h &= \text{ReLU}(W_0^T z), \\
s &= W_1^T h, \\
M_c &= \text{Softmax}(\theta \odot s),
\end{align}
\noindent where $V_{\text{avg}}$ $\in$ $\mathbb{R}$$^{c \times 1}$ and $V_{\text{max}}$ $\in$ $\mathbb{R}$$^{c \times 1}$ represent the average-pooling and max-pooling outputs of the input vectors. 
These two vectors are fused via element-wise summation to obtain the intermediate vector $z$.
The average-pooling preserves global contextual, whereas max-pooling emphasizes the most discriminative features.
The weight matrix $W_0^T$ $\in$ $\mathbb{R}^{6 \times c}$ projects the vector $z$ into a 6-dimensional hidden space, followed by the ReLU activation function to obtain the intermediate representation $h \in \mathbb{R}^{6 \times 1}$.
Subsequently, $h$ is mapped back to a $c$-dimensional vector $s$ $\in$ $\mathbb{R}^{c \times 1}$ via the projection matrix $W_1^T$ $\in$ $\mathbb{R}^{c \times 6}$.
A masking vector $\theta$ $\in$ $\mathbb{R}$$^{c \times 1}$ is applied element-wise to suppress inactive channels before applying a Softmax function.
In practice, to ensure that inactive channels receive zero attention, we assign a large negative value (\ie, -$10^{-9}$) to the corresponding positions.
Finally, $M_c$ $\in$ $\mathbb{R}^{c \times 1}$ is the resulting channel attention map.}

\begin{table}[t]
\caption{Parameters of our MLP-based Android malware classifier}\label{para}
\resizebox{\linewidth}{!}{
\begin{tabular}{|c|c|c|c|}
\hline
\textbf{Designed Layer} & \textbf{Dimension} & \textbf{Activation Function} & \textbf{Using Dropout} \\ \hline
Input Layer             & 128                & None                         & No                    \\ \hline
Hidden Layer1           & 256                & ReLU                         & Yes                    \\ \hline
Hidden Layer2           & 128                 & ReLU                         & No                    \\ \hline
Hidden Layer3           & 64                 & ReLU                         & Yes                    \\ \hline
Hidden Layer4           & 32                 & ReLU                         & No                    \\ \hline
Hidden Layer5           & 16                 & ReLU                         & Yes   \\ \hline
Output Layer            & 1                  & Sigmoid                      & No                     \\ \hline
\end{tabular}}
\end{table}
% \vspace{-10mm} 

After that, we make element-wise multiplication
of $M_c$ and $V$ to produce attention-weighted feature vectors.
% $V^F \in \mathbb{R}^{c \times d}$.
We next merge the vectors using element-wise summation:
\begin{equation}
    y^{(0)} = \mathbf{1}^\top (M_c \cdot V)\textcolor{blue}{.}
\end{equation}

\subsubsection{Malware Classifier}
The fused feature $y^{(0)}$ $\in \mathbb{R}^{1 \times d}$ is fed into a 6-layer multi-layer perceptron (MLP) to get malware detection results.
The parameters used in the model are listed in \Cref{para}.
% Let $l$ = \{0, 1, 2, 3, 4, 5, 6\} be a layer in our model, $z^{(l)}$ be the linear transformation output of the layer $l$, $y^{l}$ be the output vector of the layer $l$, $W^{l}$ be the weights of the layer $l$, $b^{l}$ be the biases of the layer $l$. 
% $f$ be the ReLU activation function and $o$ be the Sigmoid output function.
% The equation for the feed-forward operation is shown as follows:
%  \begin{align}
%  z^{(l+1)} &= W^{(l+1)}y^{l} + b^{(l+1)} \\
%  y^{l+1} &= \text{ReLU}(z^{(l+1)}), \text{ where } l \neq 6 \\ 
%  y^{(6)} &= \text{Sigmoid}(z^{(6)})
% \end{align}
% The predicate label $L$ for the input vector $V$ is listed below:
% $$L = \begin{cases}
%   \text{1},\ \text{if}\ y^{(6)} \geq \text{0.5} \\
%   \text{0},\ \text{if}\ y^{(6)} < \text{0.5}
% \end{cases}
% $$
During the learning process, our model is tuned to minimize the value of the loss function, \ie, the cross-entropy function.
% The loss function is described below:
% $$J(L, L') = -\sum_{i=1}^{2}L^{(i)}logL'^{(i)}+(1-L^{(i)}log(1-L'^{(i)}))$$ 
% \noindent, where $L$ denotes the label of the input vector, and $L'$ denotes the predicted label, and $i$ indicates the real categories of the input apps, \ie, malware and benignware.
Moreover, to avoid the overfitting problem, Dropout regularization is applied for skipping some units randomly while the model is trained.
%!TEX root = bare_jrnl.tex

\section{Experimental Evaluation}

To evaluate the effectiveness of \oursystem, we seek to answer the following three questions:

\begin{itemize}
    \item \textbf{RQ1:} Does \oursystem detect Android malware better than representative tools?
    Whether \textit{flow2vec} facilitates precise malware detection compared to other typical techniques?
    How does \oursystem perform in terms of app evolution?
    What’s the time cost of \oursystem?
    \item \textbf{RQ2:} Is it necessary to perform the flow-context-aware graph refinement? 
    How effective is \textit{flow2vec} in representing app behaviors with different usage of meta-paths?
    % What is the impact on detection performance when the joint sampling over multiple meta-paths is replaced by the direct addition of different vertor representations?
    \item \textbf{RQ3:} How effective is the model for fusing the semantics of heterogeneous flows used in \oursystem for Android malware detection?
    What is the stability of the model?
    \textcolor{black}{What is the optimal depth of our MLP-based classifier?}
\end{itemize}

\subsection{Experimental Setup}\label{chap:setup}
\subsubsection{Implementation} We \textcolor{black}{implemented} a prototype of \oursystem.
We \textcolor{black}{used} the off-the-shelf tools~\cite{DBLP:conf/pldi/ArztRFBBKTOM14,DBLP:conf/icse/0029BBKTARBOM15,meng2019appscalpel} to get flow-related analysis results from apps.
We set the timeout for each tool in analyzing an app as 20 minutes.
We \textcolor{black}{leveraged} Deep Graph Library to build and refine the HIN, and \textcolor{black}{achieved} the meta-path-group-guided random walk.
We finally \textcolor{black}{implemented} the channel-attention-based DNN classifier by PyTorch.
Our MLP-based classifier \textcolor{black}{was} modeled with the dropout rate of 0.5 and the learning rate of 0.001, both of which are commonly-used in typical DNN-based models.
The experiments \textcolor{black}{were} conducted on a machine with AMD Ryzen 5 7600X 6-Core Processor 4.70 GHz CPU, 128GB  memory, Ubuntu 22.04 LTS (64bit) and NVIDIA GeForce RTX 3080 Ti (12GB) GPU.

\begin{table}
\centering
\caption{Number of Apps in different categories of our dataset}\label{tab:dataset}
\resizebox{\linewidth}{!}{
\begin{tabular}{|l|l|l|l|l|l|l|l|c|}
\hline
\textbf{Category} & \textbf{2017} & \textbf{2018} & \textbf{2019} & \textbf{2020} & \textbf{2021} & \textbf{2022} & \textbf{2023} & \textbf{2024} \\
\hline
Benignware & 1132 & 3126 & 2631 & 2929 & 2314 & 3125 & 1204 & 206 \\
\hline
Malware & 1022 & 1994 & 2448 & 2096 & 3195 & 1808 & 1893 & 178 \\
\hline
% Sum & 2154 & 5120 & 5079 & 5025 & 5509 & 4933 & 3097 & 384 \\
\end{tabular}}
\end{table}

\subsubsection{Dataset}\label{chap:dataset} To ensure authenticity and reliability of our
statistics and the experimental results, as shown in \Cref{tab:dataset}, we randomly collect 31,301 real-world samples spanning multiple years from AndroZoo~\cite{allix2016androzoo}.
The average size of each sample is 19.75 MB, where a maximum size of 333.04 MB and a minimum size of 4 KB. A sample is regarded as malicious if it is flagged by more than 2 antivirus engines, and a sample is regarded as benign if no engine reports it~\textcolor{black}{\cite{zhu2020measuring}}.
\textcolor{black}{
The relatively permissive labeling threshold is intended to ensure sufficient coverage of potentially malware samples.}

To extract various flow-based features from the apps above, we rent 12 cloud servers and run our self-developed extractor for over 3 months.
Note that all the samples are analyzed by the aforementioned analysis tools without errors and interruptions.
From malware samples, we obtain a total of 3,926,619 sensitive data-flow paths, 218,021 suspicious condition structures, and 157,153 ICC links; from benignware samples, we get a total of 13,905,513 sensitive data-flow paths, 108,472 suspicious condition structures, and 257,766 ICC links.
All extractions are used to build the HIN.

\subsubsection{Baselines} To demonstrate the effectiveness of our work, we select four representative and related Android malware detection tools with different key techniques:
\begin{itemize}
    \item HinDroid~\cite{hou2017hindroid} models apps, APIs, and their relations as a HIN, and computes app similarities via multi-kernel learning over semantic meta-paths.
    Based on the open-source project~\cite{HinDroidProject}, we run HinDroid with three improvements: API reduction, node2vec~\cite{grover2016node2vec}, and metapath2vec.
    \textcolor{black}{Note that HinDroid with node2vec uses a DNN-based classifier, while HinDroid with the other two improvements uses a SVM-based classifier.}
    \item AppPoet~\cite{zhao2025apppoet} is a LLM-assisted Android malware detector. It firstly extracts string-type features from multiple views. Next, it generates comprehensive cross-view descriptions through multi-view prompt engineering. Finally, these descriptions are embedded and used to train a DNN model for detection.
    \textcolor{black}{We confirm with the authors of AppPoet that their method uses GPT-4o, and accordingly, we also adopt GPT-4o to ensure a fair comparison.}
    
    \item Drebin~\cite{arp2014drebin} is a typical framework that detects an app by collecting a wide range of features, \eg, used hardware, API calls and permissions from \textit{AndroidManifest.xml} and \textit{.dex} code, and then trains a SVM-based classifier.
    \item MaMaDroid~\cite{onwuzurike2019mamadroid} abstracts each API on call graph to build a first-order Markov chain, and then uses transition probabilities as features.
    Following prior findings~\cite{gao2023obfuscation}, we adopt the package mode for better performance, and apply Random Forests for malware classification.
\end{itemize}

\begin{table}
\centering
\caption{Comparison of \oursystem and other selected baselines on our whole dataset, where HinDroid-n2v = HinDroid with node2vec, HinDroid-m2v = HinDroid with metapath2vec, HinDroid-rec = HinDroid with API reduction} \label{tab:overall}
\resizebox{0.9\linewidth}{!}{
\begin{tabular}{|l|c|c|c|c|} \hline     
\textbf{Technique} & \textbf{Accuracy} & \textbf{Precision} & \textbf{Recall} & \textbf{F$_1$-score} \\ \hline 
\rowcolor{gray!20}
\multicolumn{5}{|c|}{\textbf{Detection Tool}} \\
\hline
HinDroid-n2v & 93.42\% & 95.85\% & 90.98\% & 0.9336 \\ \hline 
HinDroid-m2v & 87.33\% & 88.95\% & 85.74\% & 0.8731 \\ \hline 
HinDroid-rec & 95.33\% & 95.26\% & 95.57\% & 0.9542 \\ \hline 
MaMaDroid & 91.14\% & 94.96\% & 91.85\% & 0.9337 \\ \hline
Drebin & 93.97\%& 94.30\% & 92.40\% & 0.9333 \\ \hline 
AppPoet& 95.81\%	& 91.28\%	& 94.99\%	&0.9310\\ \hline 
\rowcolor{gray!20}
\multicolumn{5}{|c|}{\textbf{Graph Embedding Model}} \\
\hline
DeepWalk & 83.91\%& 88.80\% & 75.04\% & 0.8134 \\ \hline 
LINE & 77.85\% & 80.37\% & 73.33\% & 0.7669 \\ \hline 
node2vec & 90.78\% & 91.17\% & 88.54\% & 0.8983 \\ \hline 
metapath2vec & 93.81\% & \textcolor{black}{\textbf{99.76\%}} & 91.33\% & 0.9536 \\ \hline 
HAN & 94.51\% & 97.22\% & 92.01\% & 0.9455 \\ \hline 
\textcolor{black}{HGT} & \textcolor{black}{94.54\%} & \textcolor{black}{97.33\%} & \textcolor{black}{90.35\%} & \textcolor{black}{0.9371} \\
\hline
\textcolor{black}{DHN} & \textcolor{black}{95.08\%} & \textcolor{black}{98.31\%} & \textcolor{black}{90.65\%} & \textcolor{black}{0.9433} \\
\hline
\rowcolor{gray!20}
MalFlows & \textbf{98.34\%} & 98.98\% & \textbf{98.64\%} & \textbf{0.9881} \\ \hline
\end{tabular}}
\end{table}

Note that we attempt to find other HIN-based tools~\cite{ye2019out,hou2021disentangled,hei2021hawk,shen2024ghgdroid} but they are either not publicly available or their code fails to run properly.
Afterwards, to evaluate the performance of \textit{flow2vec}, we select 7 generic models commonly used in some well-known malware detection systems as follows:
\begin{itemize}
    \item HAN~\cite{wang2019heterogeneous} is a heterogeneous graph representation learning model that leverages predefined meta-paths and hierarchical attention mechanisms for node embedding.
    \item metapath2vec is a heterogeneous graph representation learning model that leverages predefined meta-paths and skip-gram-based random walks for node embedding.
    \item LINE~\cite{tang2015line} is a network representation learning model that preserves both first-order and second-order proximities for node embedding.
    \item node2vec is a network representation learning model that utilizes biased random walks and the skip-gram model for node embedding.
    \item DeepWalk~\cite{perozzi2014deepwalk} is a network representation learning model that utilizes truncated random walks and the skip-gram model for node embedding.
    \item \textcolor{black}{HGT~\cite{hu2020heterogeneous} is a representative heterogeneous graph neural network that captures rich semantic information through type-aware attention mechanisms.}
    \item \textcolor{black}{DHN~\cite{shi2023distance} incorporates distance encoding to enhance the expressiveness of node embeddings by capturing both semantic and structural information.}
\end{itemize}

\subsubsection{Evaluation Metrics}
We define true positives as correctly classified malware, false positives as misclassified benignware, true negatives as correctly classified benignware, and false negatives as misclassified malware.
We evaluate with four metrics, including Precision, Recall, Accuracy and F$_{1}$-score.

\subsection{RQ1: Detection on Real-World Samples}
\subsubsection{Overall Effectiveness}\label{chap:overallMalware}
To validate the effectiveness of \oursystem in malware detection, we randomly select 80\% of apps from our dataset to form the training set, and test on the rest of the apps. The process above is run for 5 times, each time using a different subset for testing, and we calculate the average for each metric as the results.
Furthermore, we train our model with a fixed 100 epochs.
\textcolor{black}{In each epoch, we set the maximum walk length to 30, meaning each walk consists of 30 complete meta-path instances at most.
This setting is applied to each meta-path group used in \textit{flow2vec}.}

The experimental results are shown at Lines 2-8 and Line 17 of \Cref{tab:overall}.
From the table, we can see that \oursystem outperforms all other selected detection tools in Recall, Accuracy and F$_1$-score.
Specifically, for the three improvements of HinDroid, HinDroid-rec outperforms HinDroid-n2v and HinDroid-m2v, which indicates that retaining HinDroid's representation method while optimizing API selection yields better results than replacing its representation method.
Furthermore, the F$_1$-score of HinDroid-rec is better than that of MaMaDroid, Drebin, and AppPoet.
This demonstrates the value of using the HIN to detect Android malware.
Compared to HinDroid-rec, \oursystem provides a comprehensive assessment of apps by complementarily fusing the semantics from the views of data flows, control flows, and ICCs, rather than relying solely on the relations of API calls.
Similarly, MaMaDroid focuses on the sequences of API calls, which only reflects a portion of the semantics extracted from the three views used in \oursystem.
Moreover, the features used by AppPoet and Drebin are organized as sets of strings, whereas \oursystem extracts high-order features from the HIN that contains rich relations based on different types of flows, enabling the expression of more structural and semantic information.
 
\begin{table}[t]
\centering
\caption{Comparison of \oursystem and other tools in the app evolution, where HinDroid corresponds to HinDroid-rec in \Cref{tab:overall} as it obtains the best results of the three improvements}\label{tab:evo} 
\resizebox{0.85\linewidth}{!}{
\begin{tabular}{|l|c|c|c|c|} \hline     
\multirow{2}{*}{\textbf{Tool}} & \multicolumn{4}{c|}{\textbf{App Evolution}} \\ \cline{2-5} 
 & \textbf{AUT(a)} & \textbf{AUT(p)} & \textbf{AUT(r)} & \textbf{AUT(f$_1$)} \\ \hline 
HinDroid & 0.8871 & 0.8860 & 0.8841 & 0.8849 
\\ \hline 
MaMaDroid& 0.8515 & 0.8363 & 0.8370 & 0.8364 
\\ \hline 
Drebin& 0.8793 & 0.8715 & 0.8492 & 0.8599 
\\ \hline 
AppPoet& 0.9153 & 0.9199 & 0.9187 &0.9185 
\\ \hline 
MalFlows& \textbf{0.9369} & \textbf{0.9275} & \textbf{0.9446} & \textbf{0.9358} 
\\ \hline
\end{tabular}}
\end{table}

\subsubsection{Effectiveness of flow2vec}\label{chap:flow2vec1}
To verify the effectiveness of the HIN embedding in \oursystem, we compare \textit{flow2vec} with the representative baseline models described above.
Specifically, we replace the graph embedding module in \oursystem, while keeping the dataset used, as well as the training and testing processes, identical to those described in \Cref{chap:overallMalware}.

The experimental results are listed at Lines 9-16 and Line 17 of \Cref{tab:overall}, showing that the HIN embedding method used in \oursystem outperforms all other models across comprehensive metrics.
Specifically, metapath2vec, HAN, \textcolor{black}{HGT and DHN} perform better than DeepWalk, LINE, and node2vec.
This implies that, compared with modelling Android app behaviors with homogeneous graphs, using heterogeneous graphs can describe app behaviors more precisely and enhance the distinguishability between malware and benignware.
\textcolor{black}{Furthermore, compared with a range of heterogeneous graph embedding models, MalFlows consistently achieves superior performance in terms of Accuracy, Recall, and F$_1$-score.}
In \oursystem, the meta-path-group-guided random walk of \textit{flow2vec} jointly samples the HIN across multiple meta-paths under specified views, which helps capture structural and semantic relations in the HIN.
Therefore, the learned vector embeddings can profile app behaviors comprehensively and precisely, leading to the best detection performance among all evaluated schemes.
\textcolor{black}{Although metapath2vec achieves the highest precision (99.76\%) in our experiments, its relatively lower recall suggests potential limitations in capturing diverse semantics of malicious app behaviors.}
Note that all the heterogeneous graph embedding models are evaluated on our refined HIN.
Their performance would likely deteriorate if applied to the standard HIN.

\begin{figure}[t]
    \centering
    \begin{minipage}[t]{0.498\columnwidth}
        \includegraphics[width=\textwidth]{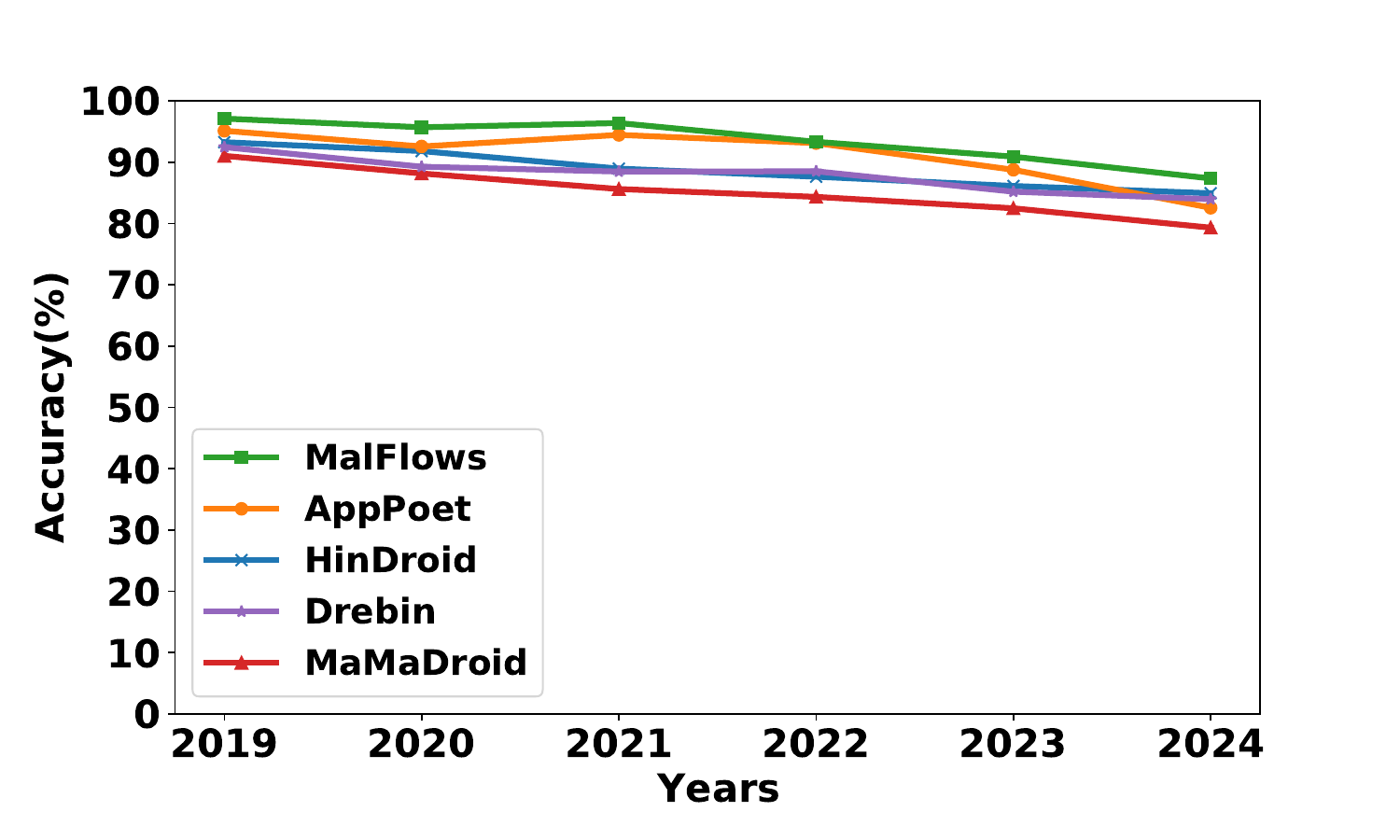}
    \end{minipage}%
    \hspace{0.1em}%
    \begin{minipage}[t]{0.498\columnwidth}
        \includegraphics[width=\textwidth]{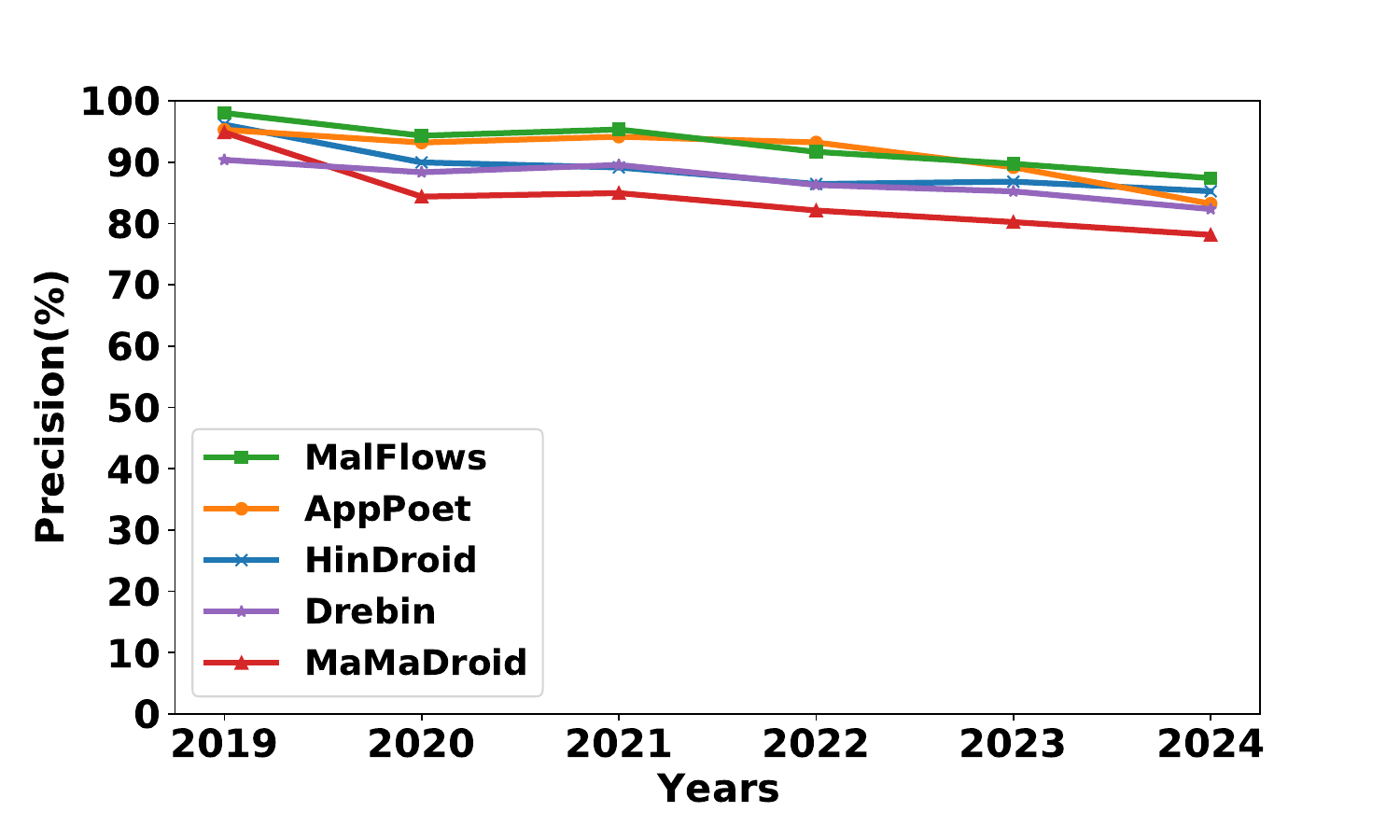}
    \end{minipage}
    
    \vspace{0.1em}
    
    \begin{minipage}[t]{0.498\columnwidth}
        \includegraphics[width=\textwidth]{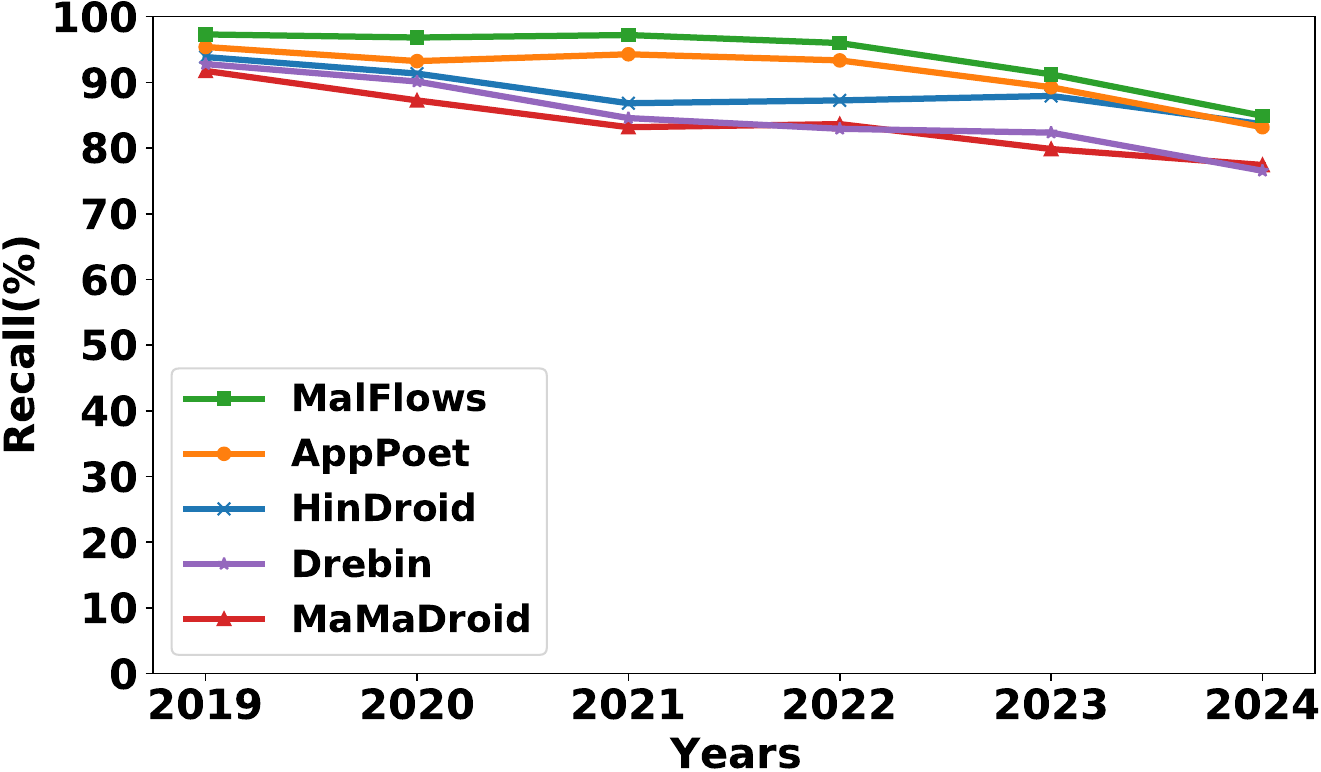}
    \end{minipage}%
    \hspace{0.1em}%
    \begin{minipage}[t]{0.498\columnwidth}
        \includegraphics[width=\textwidth]{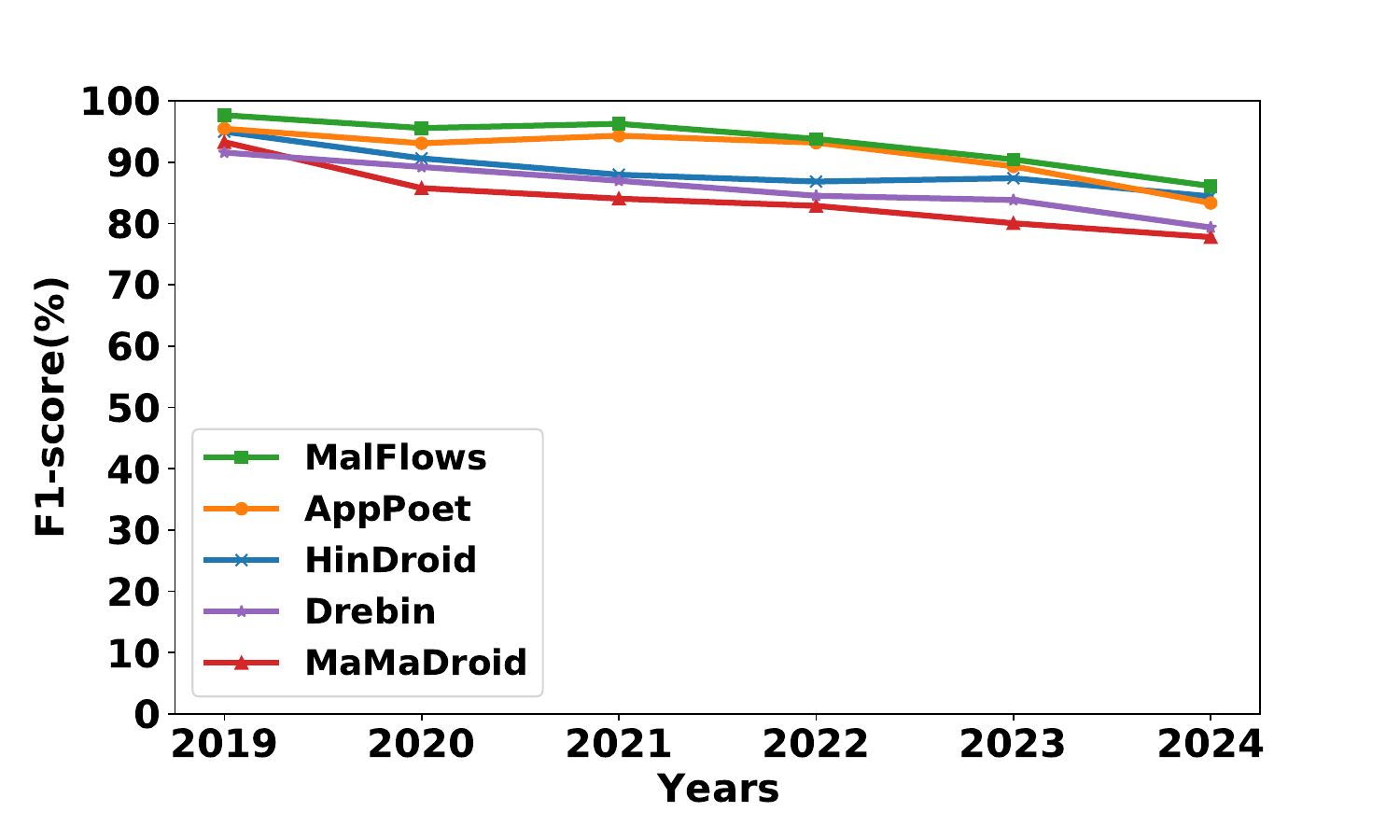}
    \end{minipage}
    \caption{\textcolor{black}{Effectiveness metrics over 6 test slots of total 72 months.}}
    \label{fig:evoTrend}
\end{figure}

\subsubsection{Adaptation for App Evolution}
Machine learning-based Android malware detectors face a problem of aging as the app evolution due to the update of Android version~\cite{zhang2020enhancing}.
To construct the required experimental environment, we use the samples from 2018 as the training set and the samples from each year between 2019 and 2024 to form six separate testing sets.
Note that the samples from 2017 are not included due to their insufficient size.
The appearance timestamps of the apps in the training set are earlier than the appearance timestamps of the apps in the testing set, so we experiment to figure out whether a malware classifier can identify evolved Android
malware.
We use Area Under Time (AUT) to measure a classifier’s robustness to time decay~\cite{pendlebury2019tesseract}:
\begin{equation}
AUT(f,N)= \frac{1}{N-1} \sum_{k=1}^{N-1} \frac{f(k+1) + f(k)}{2}\textcolor{blue}{,}
\end{equation}
where $f$ is our evaluation metric, $N$ is the number of test slots, and $f(k)$ is the evaluation metric generated at time $k$.
N is set as 12 months in our experiment and thereby omitted from AUT expressions in subsequent sections.
An AUT metric that is closer to 1 means better performance over time.

The experimental results are listed in \Cref{tab:evo}, where \oursystem outperforms all other tools in AUT(p), AUT(r), AUT(a) and AUT(f$_{1}$).
\textcolor{black}{Note that \Cref{tab:evo} includes only Android malware detection methods.
Since the app evolution scenario requires evaluating the robustness of the entire detection pipeline, it would be inappropriate to compare standalone graph representation techniques in this context.}
This superior effectiveness is attributed to the complementary nature of the three flow-related views used by \oursystem, which facilitates capturing intrinsic factors of apps.
Furthermore, it is interesting that AppPoet demonstrates the best overall performance among all other baseline tools. \textcolor{black}{Given that ChatGPT was trained on a large-scale corpus that may include or resemble some samples from the dataset used in our evaluation, there exists a possibility of training data leakage. 
Such leakage could result in more accurate or informative summaries for certain testing samples, which may partially explain the strong performance of AppPoet.}
Moreover, the results can validate that the features embedded from the content supplemented by LLMs are more stable over time than only API-related features and discrete string-based features.
However, the content in string format cannot intuitively represent the structural and semantic relations inherent in flow-related information, causing AppPoet to overlook key invariants in app evolution.
This limitation ultimately results in AppPoet performing less effectively than MalFlows.
% Therefore, AppPoet is less adaptable to app evolution compared to MalFlows.

% \begin{table}
% \centering
% \caption{Effectiveness of \oursystem in Android malware detection without the graph refinement}\label{tab:graphRefiment}
% \resizebox{0.7\linewidth}{!}{
% \begin{tabular}{|c|c|c|c|}
% \hline
%  \textbf{Accuracy} & \textbf{Precision} & \textbf{Recall} & \textbf{F$_1$-score} \\
% \hline
% 84.30\% & 96.54\% & 79.72\% & 87.33\% \\
% \hline
% \end{tabular}}
% \end{table}

\begin{table}
\centering
\caption{\textcolor{black}{Average time cost comparison of \oursystem and baseline techniques}}\label{tab:timecost}
\resizebox{0.8\linewidth}{!}{
\begin{tabular}{|ccc|}
\hline
\multicolumn{1}{|c|}{\textbf{Technique}} & \multicolumn{2}{c|}{\textbf{Time Cost (milliseconds)}}                \\ \hline
\rowcolor{gray!20}\multicolumn{3}{|c|}{\textbf{Detection Tool}}                                                                    \\ \hline
\multicolumn{1}{|l|}{}                   & \multicolumn{1}{c|}{\textbf{Training Time}} & \textbf{Inference Time} \\ \hline
\multicolumn{1}{|c|}{HinDroid-n2v}       & \multicolumn{1}{c|}{0.03}                   & 0.01                  \\ \hline
\multicolumn{1}{|c|}{HinDroid-m2v}       & \multicolumn{1}{c|}{4.49}                   & 0.06                    \\ \hline
\multicolumn{1}{|c|}{HinDroid-rec}       & \multicolumn{1}{c|}{2.83}                   & 0.03                    \\ \hline
\multicolumn{1}{|c|}{MaMaDroid}          & \multicolumn{1}{c|}{0.12}                   & 0.11                    \\ \hline
\multicolumn{1}{|c|}{Drebin}             & \multicolumn{1}{c|}{7.22}                   & 0.02                    \\ \hline
\multicolumn{1}{|c|}{AppPoet}            & \multicolumn{1}{c|}{0.54}                   & 0.40                    \\ \hline
\multicolumn{1}{|c|}{MalFlows}           & \multicolumn{1}{c|}{0.09}                   & 0.02                    \\ \hline
\rowcolor{gray!20}\multicolumn{3}{|c|}{\textbf{Graph Embedding Model}}                                                             \\ \hline
\multicolumn{1}{|c|}{DeepWalk}           & \multicolumn{2}{c|}{1273.24}                                                 \\ \hline
\multicolumn{1}{|c|}{LINE}               & \multicolumn{2}{c|}{401.95}                                                 \\ \hline
\multicolumn{1}{|c|}{node2vec}           & \multicolumn{2}{c|}{1730.73}                                                 \\ \hline
\multicolumn{1}{|c|}{metapath2vec}       & \multicolumn{2}{c|}{1284.40}                                                 \\ \hline
\multicolumn{1}{|c|}{HAN}                & \multicolumn{2}{c|}{1236.83}                                                 \\ \hline
\multicolumn{1}{|c|}{HGT}                & \multicolumn{2}{c|}{448.58}                                                 \\ \hline
\multicolumn{1}{|c|}{DHN}                & \multicolumn{2}{c|}{4615.81}                                                 \\ \hline
\multicolumn{1}{|c|}{\textit{flow2vec}}           & \multicolumn{2}{c|}{1597.19}                                                 \\ \hline
\end{tabular}}
\end{table}

We then plot the change trend of the evaluation metrics for each selected tool over time in \Cref{fig:evoTrend}.
Overall, the metrics of all the tools drop to varying degrees over time.
Among them, \oursystem exhibits the slowest decline in the four AUT metrics, indicating
its robustness against the aging process.
In contrast, the effectiveness of HinDroid, MaMaDroid, Drebin, and AppPoet deteriorates after 5 years, with all the metrics falling below 0.9.
Therefore, we suggest that a detection model should ideally be retrained with newer app data after 4 years of use in practice.
The suggestion is less strict than the one obtained by LensDroid~\cite{meng2025detecting}.
Another interesting finding is that the metrics of AppPoet remain stable or even improve during 2020–2022, followed by a gradual decline.
\textcolor{black}{This trend suggests that the dataset from 2020–2022 used in our evaluation may overlap with the training data of ChatGPT, potentially contributing to the improved performance during that period. Moreover, these results imply that the feature descriptions generated by LLMs possess a certain degree of robustness against app evolution.}

\subsubsection{Time Cost}
\textcolor{black}{We conduct corresponding experiments to assess the computational efficiency of our work, and the results are presented in \Cref{tab:timecost}.}

\textcolor{black}{For detection tools, the classifier in \oursystem demonstrates excellent efficiency, with an average training time of 0.09 ms and inference time of 0.02 ms per sample.
It outperforms traditional classifiers used in Drebin, HinDroid-m2v, MaMaDroid, and HinDroid-rec, and is also faster than AppPoet.
Although HinDroid-n2v is slightly faster, it adopts a simpler embedding strategy, which leads to lower detection effectiveness, as shown in \Cref{tab:overall}.
For graph embedding models, \textit{flow2vec} incurs a moderate time cost (1597.19 ms) compared to other graph embedding models. 
Although it is slower than HAN (1236.83 ms) and some lightweight models like metapath2vec (1284.40 ms), it remains more efficient than node2vec (1730.73 ms) and DHN (4615.81 ms).
Furthermore, HIN construction in \oursystem is lightweight (0.23 ms).}

\textcolor{black}{Notably, although the meta-path-group–guided sampling strategy in \textit{flow2vec} introduces additional computational overhead, \oursystem still achieves the best detection performance with acceptable time cost, indicating a favorable trade-off between effectiveness and efficiency.}

% \oursystem consists of two stages.
% The HIN embedding stage is relatively time-consuming.
% The maximum runtime for a single view on our dataset is approximately 47 minutes.
% However, this stage is conducted offline and only once per dataset.
% Moreover, its efficiency can be improved through parallelization or enhanced hardware configurations.
% In contrast, the detection stage is highly efficient, making the system suitable for real-time analysis.
% \oursystem takes roughly 2.50 milliseconds on average to train a batch of 128, resulting in 1.04 seconds for an epoch over the whole dataset.
% The prediction time on average is 0.11 milliseconds.

\begin{table}[t]
\centering
\caption{\textcolor{black}{Structural changes in our constructed HIN before and after applying the graph refinement}}\label{tab:diff}
\resizebox{0.8\linewidth}{!}{
\begin{tabular}{|lccc|}
\hline
\multicolumn{1}{|l|}{\textbf{Type}} & \multicolumn{1}{c|}{\textbf{Before\#}} & \multicolumn{1}{c|}{\textbf{After\#}} & \textbf{Diff} \\ \hline
\rowcolor{gray!20}
\multicolumn{4}{|c|}{\textbf{Node}}                                                                                                     \\ \hline
\multicolumn{1}{|l|}{App}           & \multicolumn{1}{c|}{31,351}             & \multicolumn{1}{c|}{31,351}            & 0                \\ \hline
\multicolumn{1}{|l|}{Action}        & \multicolumn{1}{c|}{23,416}             & \multicolumn{1}{c|}{23,416}            & 0                \\ \hline
\multicolumn{1}{|l|}{Comp}          & \multicolumn{1}{c|}{93,301}             & \multicolumn{1}{c|}{262,699}           & +169,398          \\ \hline
\multicolumn{1}{|l|}{API}           & \multicolumn{1}{c|}{39,330}             & \multicolumn{1}{c|}{684,628}           & +645,298          \\ \hline
\multicolumn{1}{|l|}{Cond}          & \multicolumn{1}{c|}{10}                & \multicolumn{1}{c|}{22,453}            & +22,443           \\ \hline
\rowcolor{gray!20}
\multicolumn{4}{|c|}{\textbf{Edge}}                                                                                                     \\ \hline
\multicolumn{1}{|l|}{App-Action}    & \multicolumn{1}{c|}{1,350,516}           & \multicolumn{1}{c|}{1,350,516}          & 0                \\ \hline
\multicolumn{1}{|l|}{App-API}       & \multicolumn{1}{c|}{671,939}            & \multicolumn{1}{c|}{671,939}           & 0                \\ \hline
\multicolumn{1}{|l|}{App-Comp}      & \multicolumn{1}{c|}{1,350,516}           & \multicolumn{1}{c|}{1,350,516}          & 0                \\ \hline
\multicolumn{1}{|l|}{App-Cond}      & \multicolumn{1}{c|}{22,453}             & \multicolumn{1}{c|}{22,453}            & 0                \\ \hline
\multicolumn{1}{|l|}{Comp-Action}   & \multicolumn{1}{c|}{1,350,516}           & \multicolumn{1}{c|}{1,350,516}          & 0                \\ \hline
\multicolumn{1}{|l|}{Cond-API}      & \multicolumn{1}{c|}{1,664}              & \multicolumn{1}{c|}{1,664}             & 0                \\ \hline
\multicolumn{1}{|l|}{API-API}       & \multicolumn{1}{c|}{242,802}            & \multicolumn{1}{c|}{242,802}           & 0                \\ \hline
\rowcolor{gray!20}
\multicolumn{4}{|c|}{\textbf{Meta-path}}  \\ \hline
\multicolumn{1}{|l|}{MPG1}   & \multicolumn{1}{c|}{15,091,000}           & \multicolumn{1}{c|}{13,260,109}          & -1,830,891                \\ \hline
\multicolumn{1}{|l|}{MPG2}   & \multicolumn{1}{c|}{23,599,472}           & \multicolumn{1}{c|}{22,446,913}          & -1,152,559                \\ \hline
\multicolumn{1}{|l|}{MPG3}   & \multicolumn{1}{c|}{11,351,000}           & \multicolumn{1}{c|}{11,060,291}          & -290,709               \\ \hline
\end{tabular}}
\end{table}

\subsubsection{Case Study} To exhibit the preformance of \oursystem, we present a malware sample named \textit{com.runbey.byy}\footnote{\texttt{MD5: 8ed90e74e8b41ecece4c35e86ebefff4}}.
The app is detected by \oursystem correctly but misclassified by all the baselines above.
To investigate the reason behind the app's classification, we extract the weights from the channel attention map during detection.
The weights corresponding to the three channels are 0.4103, 0.2543, and 0.3354 respectively.
This indicates that \oursystem adaptively emphasizes more informative views of the flows during the decision-making process,  with particular attention given to the control-flow view, which contributes most significantly to the final prediction.
Furthermore, we analyze the app from three views as follows:
(a) In the view of control flows, the app frequently performs file operations when the network-related conditions are triggered.
(b) In the view of data flows, the app repeatedly writes data to files during execution.
(c) In the view of ICCs, the app makes use of various third-party components (\eg, \textit{cn.jpush}, \textit{com.tencent}, \textit{com.sina}) and custom actions of Intents (\eg, \textit{cn.jpush.android.intent.RECEIVE\_MESSAGE}). 
Overall, the app is suspected to contain Trojan-like behavior and is capable of downloading unknown files from the network.

\subsection{RQ2: Ablation Study of flow2vec}
\subsubsection{Necessity of Graph Refinement}

\textcolor{black}{To demonstrate the contribution of the graph refinement on the structure of our HIN, we provide a quantitative analysis of its impact.
we add a comprehensive structural comparison of our HIN before and after applying the graph refinement.
The experimental configurations are consistent with those described in Section IV-B1. The comparison results are listed in \Cref{tab:diff}.
Specifically:}

\begin{itemize}
    \item \textcolor{black}{Significant increases are observed in Comp, API, and Cond node types, with 169,398, 645,298, and 22,443 nodes added respectively.
    These nodes are enriched via flow-context-aware analysis, enabling the HIN to represent more accurate and meaningful dependencies.}
    \item \textcolor{black}{The number of edges remains unchanged, indicating that the refinement primarily focuses on node augmentation without introducing redundant connections.}
    \item \textcolor{black}{To measure invalid path reduction, we perform random walks with identical configurations before and after applying Algorithm 1, remove duplicate paths, and compare the number of unique paths.
    We observe a substantial decrease in the number of meta-path instances across the three defined meta-path groups, with reductions of 1,830,891, 1,152,559, and 290,709 for MPG1, MPG2, and MPG3 respectively.
    Although random walk introduces a certain degree of sampling randomness, the consistently significant reduction in the number of sampled paths under identical experimental configurations still clearly demonstrates the effectiveness of graph refinement in eliminating invalid semantic paths.}
\end{itemize}

We further illustrate the necessity of the graph refinement based on the malware detection results below.
Except for not using the graph refinement, we perform the same experimental configurations as described in \Cref{chap:overallMalware}.

The results are listed at the third line of \Cref{tab:mps}.
We observe that the detection performance of \oursystem drops significantly across all metrics when the graph refinement is disabled, compared to when it is employed (\ie, Ours).
After further analysis, we infer that the variant of \textit{flow2vec} without graph refinement tends to sample the paths that deviate from real-world scenarios from the built HIN.
This leads to confusion in the relations between apps, which in turn negatively affects detection performance.

\begin{table}[t]
\centering
\caption{Detection performance of \textit{flow2vec} with different Settings}\label{tab:mps}
\resizebox{0.88\linewidth}{!}{
\begin{tabular}{|l|c|c|c|c|}
\hline
\textbf{Setting}  & \textbf{Accuracy} & \textbf{Precision} & \textbf{Recall} & \textbf{F$_1$-score} \\
\hline
\rowcolor{gray!20}
\multicolumn{5}{|c|}{\textbf{Usage of Graph Refinement}} \\
\hline
Without & 84.30\% & 96.54\% & 79.72\% & 0.8733 \\
\hline
\rowcolor{gray!20}
\multicolumn{5}{|c|}{\textbf{Usage of Meta-path}} \\
\hline
MP$_1$  & 91.46\% & 95.91\% & 91.63\% & 0.9372 \\
\hline
MP$_2$  & 93.81\% & 99.76\% & 91.33\% & 0.9536 \\
\hline
MP$_3$  & 91.58\% & 94.67\% & 89.43\% & 0.9197 \\
\hline
MP$_4$  &92.49\% & 94.23\% & 90.37\% & 0.9226 \\
\hline
MP$_5$  & 88.58\% & 92.81\% & 78.09\% & 0.8481 \\
\hline
MP$_6$  &90.80\% & 93.40\% & 83.61\% & 0.8824 \\
\hline
MPG$_1$  & 94.11\% & 99.31\% & 92.35\% & 0.9570 \\
\hline
MPG$_2$  & 94.23\% & 95.72\% & 91.61\% & 0.9362 \\
\hline
MPG$_3$  & 91.71\% & 92.51\% & 87.08\% & 0.8971 \\
\hline
MPG$_1$,MPG$_2$ & 95.03\% & 96.11\% & 93.07\% & 0.9456 \\
\hline
MPG$_1$,MPG$_3$  &  94.65\% & 97.57\% & 89.46\% & 0.9334 \\
\hline
MPG$_2$,MPG$_3$  & 94.54\% & 94.74\% & 92.09\% & 0.9339 \\
\hline
\rowcolor{gray!20}
\begin{tabular}[c]{@{}l@{}}Ours \end{tabular} & \textbf{98.34\%} & \textbf{98.98\%} & \textbf{98.64\%} & \textbf{0.9881} \\
% \\ (MPG$_1$,MPG$_2$,MPG$_3$)
\hline
\end{tabular}}
\end{table}

% \begin{figure}[tb]
% \centering
% \center{\includegraphics[width=\linewidth]  {../graphs/wrongPath.pdf}}
% \caption{HIN sampling on a real-world case with and without graph refinement, where the red rectangles represent the nodes on the incorrect sampling path and the definitions of other components are the same as before.}
% \label{wrongPath1}
% \end{figure}

\subsubsection{Usage of Different Meta-paths}
% \subsubsection{Ablation Study on Meta-paths}
To verify the effectiveness of our meta-path groups for \textit{flow2vec}, we conduct ablation experiments using 13 different meta-path settings.
Experimental environments are the same as depicted in \Cref{chap:flow2vec1}.

The experimental results are presented in \Cref{tab:mps}, showing that using all groups of meta-paths (\ie, Ours) yields the best performance, as the meta-paths under the three views of the flows effectively capture and represent app behaviors.
% We find that using all groups of meta-paths can achieve the best results, where utilizing the meta-paths under the three views are expressive to profile app behaviors.
% In other words, 
% It means that the fusion of flow information from multiple views is beneficial for detecting Android malware.
We observe that: 
(a) MP$_2$ outperforms the other individual meta-paths, which indicates that trigger conditions can be effectively used to differentiate benignware and malware.
(b) MP$_2$, MP$_4$, and MP$_6$ perform better than MP$_1$, MP$_3$, and MP$_5$ correspondingly.
The reason is that important semantics within apps for detecting Android malware aggregated through action-oriented meta-paths are more beneficial than those aggregated through content-oriented meta-paths.
(c) The overall effectiveness of meta-path groups with our semantic fusion method is superior to that of most individual meta-paths.
Specifically, the Accuracy and F$_1$-score of MPG$_1$, MPG$_2$, and MPG$_3$ are higher than those of MP$_2$, MP$_4$, and, MP$_6$ respectively.
This indicates that, under a single view, joint sampling based on both action-oriented and content-oriented meta-paths is more effective in capturing key features of the HIN than sampling based on a single meta-path alone.
(d) With the combined use of multiple groups of meta-paths, the detection performance of our tool is further improved, particularly in terms of Accuracy.
For instance, with our semantic fusion method, the Accuracy of \{MPG$_1$, MPG$_2$\} is 95.03\%, which is higher than the Accuracy of MPG$_1$ (94.11\%) and MPG$_2$ (94.23\%).
This further demonstrates the complementary effect among different groups of meta-paths.

\subsection{RQ3: Performance of Our Fusion Model}
\subsubsection{Effectiveness on Fusing Heterogeneous Flow Semantics}
To validate if the flow semantic fusion in \oursystem facilitates Android malware detection,  we compare the effectiveness of \oursystem when using the original and alternative models.
The experimental environments are the same as described in \Cref{chap:flow2vec1}.
We construct three alternatives:
\begin{itemize}
    \item It uses metapath2vec to generate vectors based on the 6 meta-paths respectively, and then leverages the element-wise vector addition to fuse the vectors (\ie, Add-direct).
    \item It uses the meta-path-group-guided random walk to generate vectors for each view, and then leverages the element-wise vector addition to fuse the vectors (\ie, Add-hybrid).
    \item It implements the multi-view fusion by replacing channel attention as self-attention (\ie, Self-attn).
\end{itemize}

\Cref{tab:modelalab} presents the experimental results, showing that the model we used surpasses the alternatives in the effectiveness of malware detection.
Specifically, the methods of Add-direct and Add-hybrid perform better than the method of Self-attn.
As further analysis, we speculate that the feature vectors obtained from the three views of flows contain unique semantic information, all of which make important contributions to Android malware detection.
Vector addition does not alter the original feature representation, which makes it suitable for local feature fusion.
However, self-attention may disrupt local semantics.
Moreover, compared to Add-direct, Add-hybrid performs better, mainly because the meta-path-group-guided random walk can capture potential semantics within the HIN more effectively.
Channel attention can adaptively enhance the weights of important views while suppressing irrelevant ones, thereby improving feature representation and achieving precision Android malware detection.

\begin{table}
\centering
\caption{Comparison of the model we used and the alternative models in effectiveness of Android malware detection}\label{tab:modelalab}
\resizebox{0.95\linewidth}{!}{
\begin{tabular}{|l|c|c|c|c|} \hline     
\textbf{Fusion Method} & \textbf{Accuracy} & \textbf{Precision} & \textbf{Recall} & \textbf{F$_1$-score} \\ \hline
Add-direct & 92.00\% & 92.90\% & 89.31\% & 0.9105 \\ \hline
Add-hybrid & 95.26\% & 95.87\% & 96.07\% & 0.9597 \\ \hline
Self-attn & 87.22\% & 93.58\% & 74.57\% & 0.8300 \\ \hline 
Ours & \textbf{98.34\%} & \textbf{98.98\%} & \textbf{98.64\%} & \textbf{0.9881} \\ \hline
\end{tabular}}
\end{table}

\begin{figure}[t]
    \centering
    \subfigure[Filtered subset]{
    \begin{minipage}[t]{0.578\columnwidth}
        \includegraphics[width=\textwidth]{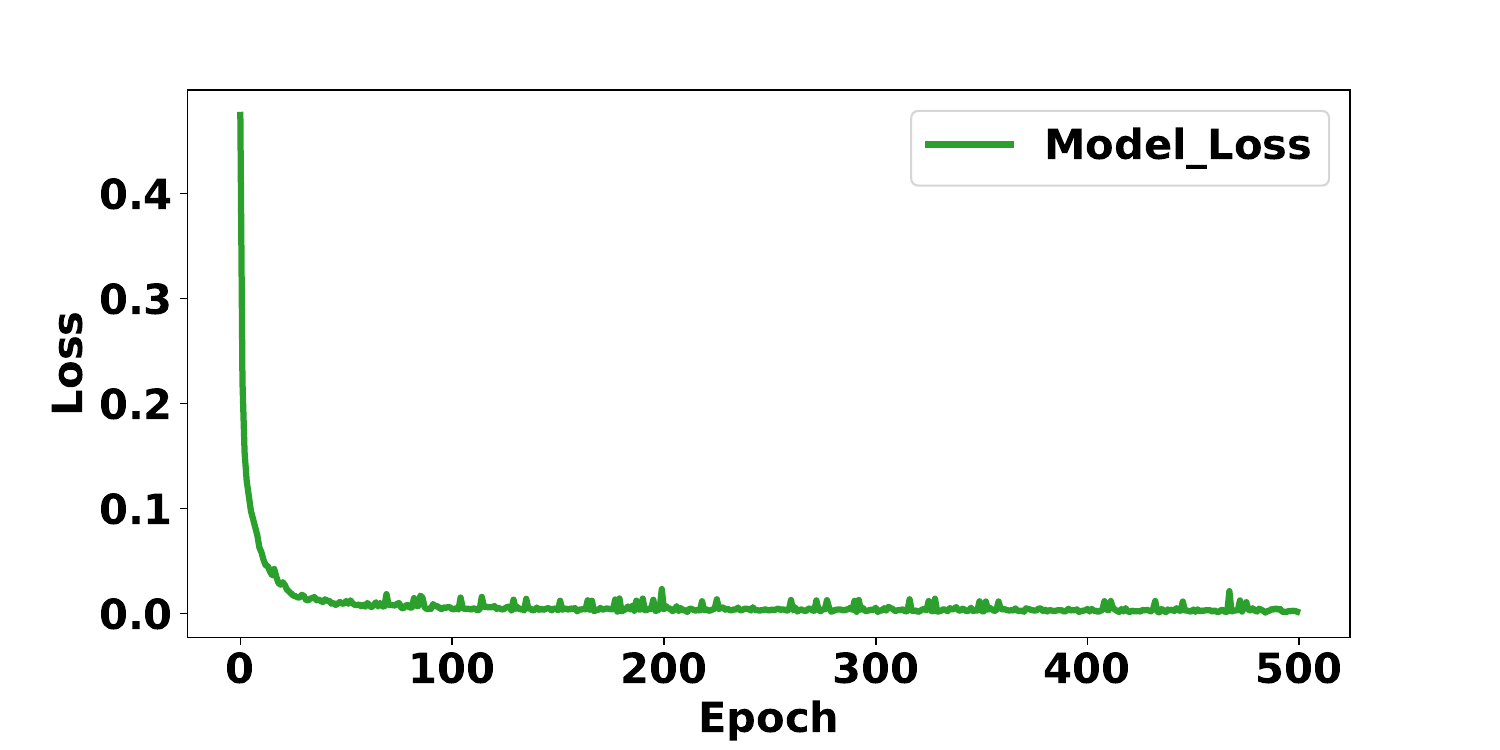}
    \end{minipage}%
    \hspace{0.1em}%
    \begin{minipage}[t]{0.416\columnwidth}
        \includegraphics[width=\textwidth]{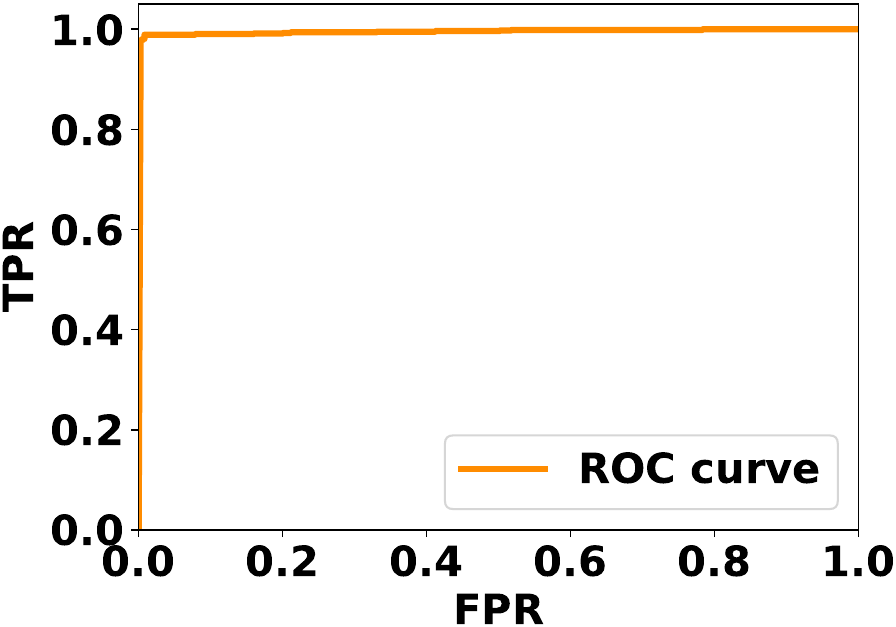}
    \end{minipage}}

    \vspace{0.1em}

    \subfigure[Whole dataset]{
    \begin{minipage}[t]{0.578\columnwidth}
        \includegraphics[width=\textwidth]{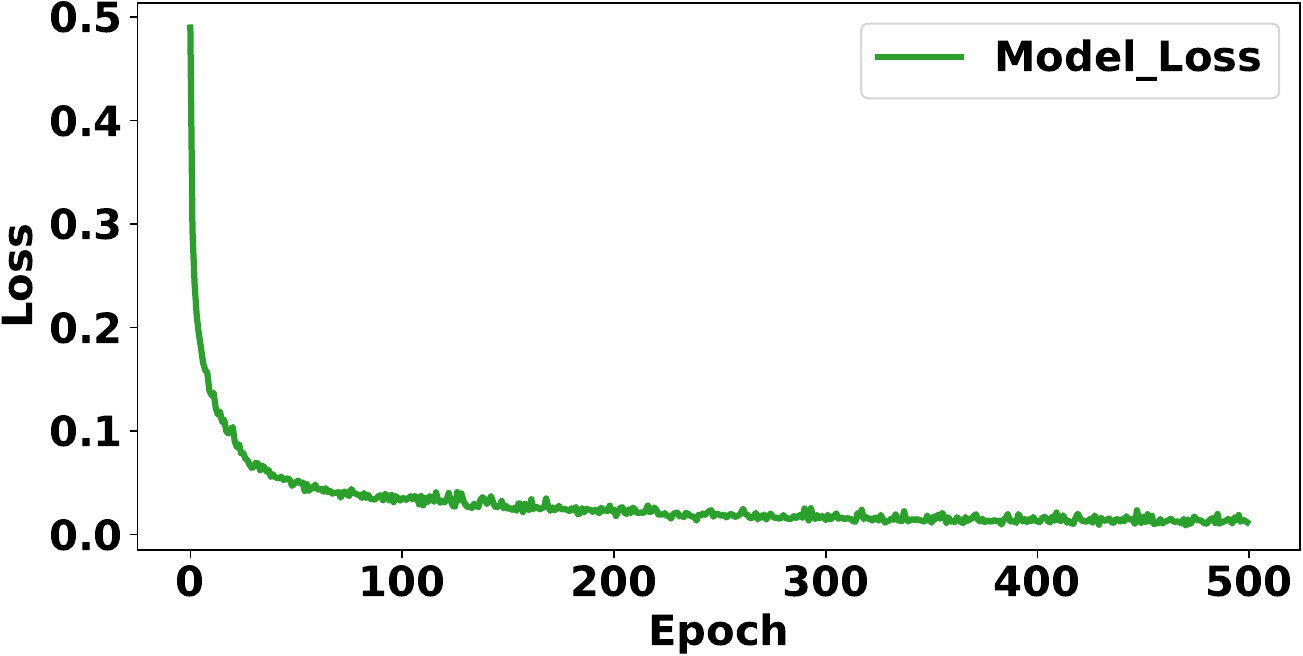}
    \end{minipage}%
    \hspace{0.1em}%
    \begin{minipage}[t]{0.416\columnwidth}
        \includegraphics[width=\textwidth]{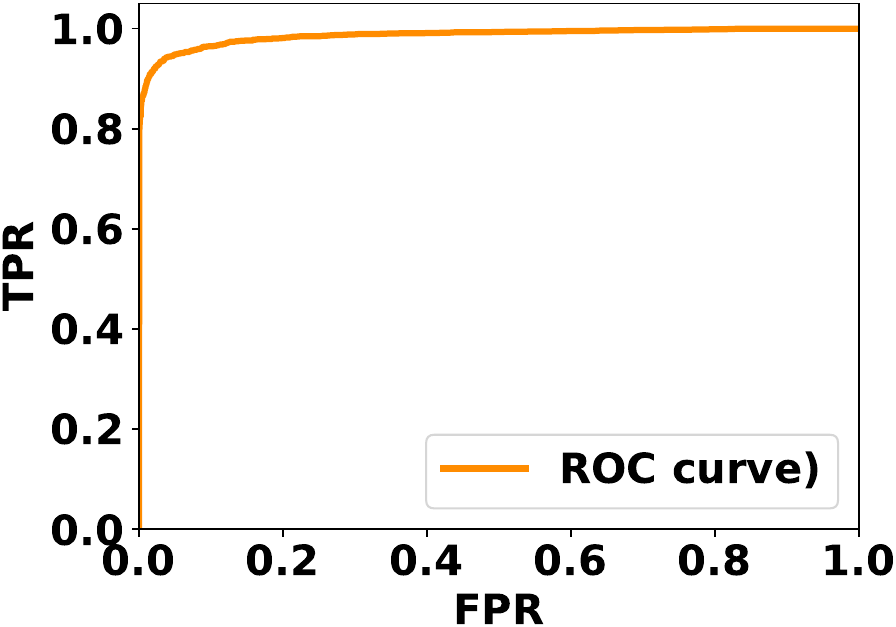}
    \end{minipage}}
    \caption{Evaluation of model stability of \oursystem.}
    \label{fig:stad}
\end{figure}

\subsubsection{Model Stability}
\textcolor{black}{To comprehensively evaluate the stability of our model, we conduct experiments on both the whole dataset (depicted in \Cref{chap:dataset}) and a filtered subset.
The filtered subset contains 12,055 apps, each with sufficient information to extract features from all three views.
This filtering strategy is adopted for the following reasons: (1) Our detection model is designed to jointly leverage all three views, and samples with missing views cannot be fully processed as intended.
(2) Allowing samples with missing views would lead to inconsistent input configurations, making it difficult to fairly evaluate the model under its intended multi-view setting.
Other experimental configurations and environments are same as the description in \Cref{chap:overallMalware}.}

\textcolor{black}{We plot \Cref{fig:stad} to demonstrate the model stability of \oursystem.
Specifically, the left two subfigures depict the training loss curves of our model on filtered subset and whole dataset respectively.
As shown in the figures, the training loss drops sharply in the early epochs and gradually converges, indicating effective model optimization. 
The final loss remains consistently low in both cases, suggesting that the model has successfully captured the underlying patterns of Android malware behaviors.
The right two subfigures show the ROC (receiver operating characteristic) curves of of the model.
The area under the curve (AUC) scores are 0.9946 with the filtered subset and 0.9854 with the whole dataset, demonstrating strong discriminative ability between benign and malicious samples.
Furthermore, the performance difference between the whole dataset and the filtered subset highlights the effectiveness of our multi-view fusion strategy, where each view contributes complementary information to robust detection.}

\subsubsection{\textcolor{black}{MLP Depth}}
\textcolor{black}{To determine the optimal depth of the MLP used in \oursystem, we conduct an ablation study by varying the number of layers from 1 to 9.
To ensure a fair comparison, all MLP variants use the same activation functions, dropout rates, and other architectural components.
\Cref{tab:layers} lists the detailed results, which demonstrate that the 6-layer MLP provides the best overall performance, particularly in terms of Accuracy (98.34\%) and F$_1$-score (0.9881).
Specifically, the overall detection performance of \oursystem improves as the MLP depth increases from 2 to 6 layers.
Beyond 6 layers, the performance slightly declines, likely due to overfitting or increased model complexity.
Therefore, we adopt the 6-layer MLP in our final design.}

\begin{table}[t]
\centering
\caption{\textcolor{black}{Detection performance of MalFlows with varying MLP depth}}\label{tab:layers}
\resizebox{0.85\linewidth}{!}{
\begin{tabular}{|l|c|c|c|c|}
\hline
\textbf{Layer} & \textbf{Accuracy} & \textbf{Precision} & \textbf{Recall} & \textbf{F$_1$-score} \\
\hline
2 & 94.85\% & 97.81\% & 95.01\%  & 0.9639 \\
\hline
3 & 96.15\% & 97.92\% & 96.65\%  & 0.9728 \\
\hline
4 & 97.08\% & 98.11\% & 97.68\%  & 0.9789 \\
\hline
5 & 97.54\% & \textbf{98.99\%} & 97.47\%  & 0.9823 \\
\hline
6  & \textbf{98.34\%} & 98.98\% & \textbf{98.64\%}  & \textbf{0.9881} \\
\hline
7 & 97.31\% & 98.34\% & 97.80\%  & 0.9807 \\
\hline
8 & 97.15\% & 98.77\% & 97.15\%  & 0.9795 \\
\hline
9 & 96.85\% & 98.65\% & 96.81\%  & 0.9772 \\
\hline
\end{tabular}}
\end{table}
%!TEX root = bare_jrnl.tex

\section{Discussion}

% In the following, we discuss certain limitations of \oursystem and suggest possible solutions for practical use.

\subsection{Threats for Disabling Our Work}

\textcolor{black}{Due to the inherent limitations of the widely-used static analysis tools, some flow-related information may be missed or inaccurately captured, potentially affecting the accuracy of our detection.
Specifically, our method does not rely on raw function names directly, but rather on abstract flow semantics extracted by off-the-shelf tools such as FlowDroid and IccTA.
In cases of lightweight code obfuscation (\eg, function renaming, call reordering), the tools can still extract correct flow relations~\cite{sun2023demystifying}, allowing our HIN construction and downstream processing remain unaffected. 
However, when facing complex obfuscation strategies, such as reflection or dynamic code loading, the tools are often unable to extract flow data, which naturally impacts our method. 
This is a known and common limitation across many static-analysis-based techniques.
In the future, we plan to more advanced program analysis techniques, \eg, hybrid analysis~\cite{tsutano2019jitana} and dynamic analysis~\cite{sutter2024dynamic}, to improve the precision and coverage of the collection of flow-related data.
Nevertheless, we emphasize that the core contribution of our work lies in achieving a HIN-based Android malware detection, rather than improving program analysis itself.}

As a ML-based detection technique, \oursystem is also susceptible to adversarial Android malware attacks~\cite{he2023efficient}.
Specifically, \oursystem performs malware detection based on the flow-based features extracted from multiple views.
Therefore, if the app code is modified from selected few of the views, \oursystem can still detect malware by leveraging the complementary information provided by other views.
Moreover, one of the fundamental principles of adversarial example generation is to preserve the functionality of the target app~\cite{li2023black}.
The flow-based features used by \oursystem are highly correlated with the app's functionality, which contributes to ensuring the invariance of these features.
In the future, we plan to select feature sources that are difficult to tamper with, \eg, information flows within C-level code.

Furthermore, our current approach is primarily designed for offline detection scenarios. In future work, we plan to incorporate out-of-distribution learning techniques~\cite{ye2019out,hei2021hawk} to establish a more comprehensive detection framework.

\subsection{Selection of HIN Representation Techniques}
\oursystem extends a random-walk-based HIN representation method to obtain feature vectors under each views.
It is well known that DNN-based HIN representation methods are also commonly used.
The rationale behind our technical choice lies in its accuracy and convenience in joint sampling based on multiple meta-paths.
As explained in \Cref{chap:rw}, multi-path joint sampling can capture complex associations among different apps, which is beneficial for the accurate detection of Android malware.
The meta-path-group-guided random walk approach is well suited to our needs and enables accurate capture of joint sampling results from diverse meta-path combinations.
Furthermore, the approach is an extension of metapath2vec and is easy to implement.

As a typical DNN-based method, HGT~\cite{hu2020heterogeneous} lacks semantic constraints imposed by predefined meta-paths, which may lead to the capture of irrelevant semantics and negatively affect detection performance.
Since HAN~\cite{wang2019heterogeneous} is inherently designed based on individual meta-paths, adapting it to support our multi-meta-path joint sampling task is non-trivial.
It may require redesigning the attention mechanism to simultaneously handle multiple semantic contexts.

% HAN ignores entities and their sequential order within each meta-paths, which may result in the loss of critical behavioral semantics; 
% Moreover,

\subsection{Usage of Meta-paths and Semantic Fusion Methods}

We define 6 meta-paths and divide them into 3 groups based on the views of heterogeneous flows.
Meta-paths within a group are used to describe contents and actions of apps under a given view respectively.
% The other meta-path designs can also be considered to capture richer relations between apps.
Experimental results demonstrate that our simplified meta-path design achieves effective detection while maintaining computational efficiency.
We will discover more suitable meta-paths automatically by LLMs~\cite{chen2024large}. 

\oursystem uses the channel attention to perform the semantic fusion of the heterogeneous flows.
Given the semantic independence and interrelation among different meta-path groups, this fusion method is well-suited for our work.
In the future, we plan to try more advanced alternatives~\cite{huang2025mfc}.
Nevertheless, we believe that the selection of the fusion method and the core contributions of our work are orthogonal.

\subsection{Dataset Construction}
\textcolor{black}{The use of VirusTotal-based labels over a long time span (2017–2024) introduces potential label inaccuracies, especially for recent samples, which may not have been fully analyzed by detection engines.
To address this issue, as described in \Cref{chap:dataset}, we adopt a relatively permissive threshold for obtaining labels from VirusTotal.
To further validate the robustness of our results, we exclude all samples from 2023 and 2024 and re-run the experiments described in \Cref{chap:overallMalware}.
The results show that \oursystem maintains consistently high performance, with an Accuracy of 98.01\%, Precision of 98.77\%, Recall of 98.29\%, and an F1-score of 0.9853.
These results indicate that the potential label uncertainty in the 2023–2024 data has minimal impact on the overall evaluation.
Moving forward, we plan to improve labeling quality by aggregating results from multiple platforms (\eg, VirusTotal, Hybrid Analysis~\cite{HybridAnalysis}, and Any Run~\cite{anyrun}), potentially adopting a voting-based strategy. We believe that such a strategy could help reduce label noise and improve the soundness of our dataset construction.}
%!TEX root = bare_jrnl.tex

\section{Related Work}
% Researchers have proposed varied approaches and tools for Android malware detection.
% We describe a part of works that are closely related to ours in the following.

\subsection{Flow-based App Analysis}

\subsubsection{Static Analysis Tools}
Mainstream detection views in static analysis include data flows, control flows, ICC usages.
Specifically, static data-flow tracking techniques~\cite{DBLP:conf/pldi/ArztRFBBKTOM14,wei2018amandroid} detected if there exist sensitive data leaks in apps.
MUDFLOW~\cite{DBLP:conf/icse/AvdiienkoKGZARB15} further identified malware based on usage characteristics of sensitive data.
Complex-Flows~\cite{shen2018android} leveraged app behavior along with information flows for classifying benign and malicious Android apps.
Static control-flow analysis are commonly adopted to find suspicious trigger conditions in apps~\cite{pan2017dark,samhi2022difuzer}.
The results of ICC detection can complement the tools above to enhance the overall detection coverage~\cite{DBLP:conf/icse/0029BBKTARBOM15}.
% DroidAPIMiner~\cite{aafer2013droidapiminer}, DroidSIFT~\cite{zhang2014semantics} and MaMaDroid~\cite{onwuzurike2019mamadroid} leverage the frequency of API calls, API dependency relations and API call sequences respectively to detect malware.
% Permpair~\cite{arora2019permpair} identifies malware by comparing permission pairs between malicious and normal samples.
ICCDetector~\cite{xu2016iccdetector} detected malware based on the specified ICC patterns.
% Moreover, some other signature-based schemes~\cite{DBLP:conf/icse/HuangZTWL14,yang2018enmobile,feng2014apposcopy} are proposed for covering more types of malware.
Each of the schemes has its own advantages in disclosing a certain type of malicious app behaviors.
In comparison, \oursystem unifies the strengths of given schemes for more comprehensive app examinations.

\subsubsection{Dynamic Analysis Tools}
% TaintDroid~\cite{enck2014taintdroid} performed runtime data-flow tracking for the Dalvik environment.
% AppsPlayground~\cite{rastogi2013appsplayground} uses an enhanced version of TaintDroid for dynamic data flow tracking.
TaintART~\cite{sun2016taintart} executed multi-level data-flow tracking for the ART environment.
Compared with TaintART, Malton~\cite{xue2017malton} monitored taint propagation instructions at more system layers.
% Furthermore, many techniques are proposed to reconstruct app behaviors via runtime information for finding hidden maliciousness~\cite{yan2012droidscope,yuan2017droidforensics,DBLP:conf/ndss/TamKFC15,meng2020appangio}.
In the future, the dynamic information-flow features can be utilized in modeling the HIN and identifying malicious behaviors complementarily.

\subsection{Heterogeneous Graph based Android Malware Detection}
% Many techniques perform Android malware detection based on the HIN.
HinDroid~\cite{hou2017hindroid} modeled the relations between APIs and apps with HIN and identifies malware using multi-kernel learning.
Scorpion~\cite{fan2018gotcha} proposed metagraph2vec, a new HIN embedding method, to represent and characterize sly malware.
AiDroid~\cite{ye2019out} performed the HIN in-sample node embeddings and then represents each out-of-sample app with convolutional neural networks.
Dr.HIN~\cite{hou2021disentangled} integrated domain priors generated from different views to devise an adversarial disentangler based on the HIN embeddings.
Hawk~\cite{hei2021hawk} modeled Android entities and behavioral relationships as a HIN and then identifies malware with graph attention networks.
GHGDroid~\cite{shen2024ghgdroid} proposed a graph-based approach that leverages a global heterogeneous graph built from sensitive APIs and graph neural embeddings to finish Android malware detection.

\oursystem and the previous HIN-based techniques profile Android apps from disparate perspectives.
Specifically, the previous techniques typically build heterogeneous graphs using existence-based information (\eg, APIs or files~\cite{hou2017hindroid,fan2018gotcha,hei2021hawk,shen2024ghgdroid}, apk signatures or affiliations~\cite{hou2021disentangled,ye2019out}).
In contrast, \oursystem constructs the HIN from fine-grained program analysis results (\eg, data-flow paths, control dependencies), and then fuses the feature vectors based on their contributions to the malware classification.
As a result, \oursystem is able to capture richer behavioral semantics of apps.
Moreover, \oursystem introduces a flow-context-aware graph refinement strategy to better exploit flow-related semantics, thereby enhancing the effectiveness of Android malware detection. 
In the future, \oursystem and the existing techniques can complement each other to enrich the compositions of the HIN, thereby enabling the detection of a wider variety of malware samples.

\subsection{Android Malware Detection based on Semantic Fusion}

With the rapid development of deep learning techniques, many advanced Android malware detection approaches based on multi-view fusion are proposed.
LensDroid~\cite{meng2025detecting} visualized app behaviors from three related but distinct views of behavioral sensitivities, operational contexts, and supported environments.
It then fused the corresponding semantics to facilitate Android malware detection.
AppPoet~\cite{zhao2025apppoet} designed the multi-view prompt engineering technique based on LLMs to integrate discrete static program features and identify malicious behaviors within apps.
Li et al.~\cite{li2025multimodal} combined features from both source code and
binary code modalities of apps to find their maliciousness.
CorDroid~\cite{gao2023obfuscation} developed an obfuscation-resilient Android malware analysis based on enhanced sensitive function call graphs and opcode-based markov transition matrixes.
AndroAnalyzer~\cite{gong2024sensitive} fused microscopic features extracted from abstract syntax trees and the macroscopic features extracted from sensitive function call graphs for Android malware detection.
Qiu et al.~\cite{qiu2022cyber} extracted features from source code, API callgraphs, and Smali opcode to find Android malware.
Kim et al.~\cite{kim2018multimodal} extracted both existence-based and similarity-based features from Android apps, and then fed the resulting vectors into a multimodal deep neural network to identify malware.

Previous studies have primarily focus on fusing app behavior information from different modalities or views that lack explicit semantic correlations. 
For example, CorDroid utilizes callgraphs and opcodes, while LensDroid combines opcodes with the content of \textit{.so} files.
As a result, much of the effort has been devoted to feature extraction and semantic alignment across heterogeneous sources.
In contrast, the heterogeneous flow information we incorporate, \eg, function call sequences and control dependencies, naturally exhibits contextual semantic relations.
This makes it more suitable for direct semantic fusion and enables more effective characterization and representation of semantics of app behaviors, thereby providing more accurate support for malicious behavior identification.

\section{Conclusion}
We propose and implement \oursystem, a novel technique that detects Android malware by fusing heterogeneous flow semantics.
Specifically, we adopt a HIN to model the rich semantics across the views derived from the three types of flow-related information. 
We further propose \textit{flow2vec}, a new HIN embedding technique that jointly leverages multiple meta-paths with contextual constraints to learn accurate app representations.
We finally feed the representations into a channel-attention-based DNN classifier for detecting malware.
Our experiments on 31,301 apps show that \oursystem achieves an Accuracy of 98.34\% and a F$_1$-score of 0.9881 in the detection, outperforming the selected baseline techniques.
We also validate the effectiveness of \textit{flow2vec} and demonstrate that the channel-attention-based semantic fusion model for different types of program flows can enhance malware detection.

\bibliographystyle{IEEEtran}
\bibliography{IEEEabrv, tex/bare_jrnl}

\ifCLASSOPTIONcaptionsoff
  \newpage
\fi

% that's all folks
\end{document}